# Constraints on the Active and Sterile Neutrino Masses from Beta-Ray Spectra: Past, Present and Future[1]

Otokar Dragoun[2], Drahoslav Vénos[3]
*Nuclear Physics Institute of the ASCR, CZ-25068 Řež, Czech Republic*

**Abstract:** Although neutrinos are probably the most abundant fermions of the universe their mass is not yet known. Oscillation experiments have proven that at least one of the neutrino mass states has $m_i > 0.05$ eV while various interpretations of cosmological observations yielded an upper limit for the sum of neutrino masses $\sum m_i <$ (0.14 – 1.7) eV. The searches for the yet unobserved 0νββ decay result in an effective neutrino mass $m_{\beta\beta} <$ (0.2 – 0.7) eV. The analyses of measured tritium β-spectra provide an upper limit for the effective electron neutrino mass $m(\nu_e) < 2$ eV. In this review, we summarize the experience of two generations of β-ray spectroscopists who improved the upper limit of $m(\nu_e)$ by three orders of magnitude. We describe important steps in the development of radioactive sources and electron spectrometers, and recapitulate the lessons from now-disproved claims for the neutrino mass of 30 eV and the 17 keV neutrino with an admixture larger than 0.03 %. We also pay attention to new experimental approaches and searches for hypothetical sterile neutrinos.

**Keywords:** Active neutrino, Beta-ray spectrum, Beta-spectrometer, Neutrino, Neutrino mass, Sterile neutrino.

## 1. INTRODUCTION

β-ray spectroscopy is a powerful tool of neutrino physics. An unexpected shape of measured β-spectra led Pauli to the concept of a new fundamental particle, the neutrino (in addition to the then known electron and proton), saving the law of energy conservation in nuclear β-decay [1]. The Fermi theory of β-decay, incorporating the Pauli neutrino and assuming a new type of particle interaction (the weak interaction), predicted the β-spectrum shape that agreed with the experiment [2]. This was a strong, though indirect, argument for the existence of the neutrino. Later on, three kinds (or flavors) of neutrinos were discovered, the electron neutrino $\nu_e$ [3], the muon neutrino $\nu_\mu$ [4] and the tau neutrino $\nu_\tau$ [5]. Of course, properties of $\nu_\mu$ and $\nu_\tau$ cannot be examined by β-spectroscopy since the β-decay energies are not sufficient to create charged μ or τ leptons born together with these neutrinos in the weak interactions.

When neutrino oscillation became a real possibility, Shrock [6] pointed out that the individual neutrino mass eigenstates should manifest themselves as irregularities (kinks) in the otherwise continuous β-spectra. This warning initiated numerous searches for an admixture of heavier neutrinos with masses $m_i$ between $10^2$ and $10^5$ eV, as allowed by the available β-decay energies (see e.g. [7]). However, up to now, β-spectroscopists have not found any kinks in their β-spectra that would correspond to those individual neutrino mass states. The upper limits for the admixture of heavier neutrinos are below 1% in most cases. It follows from oscillation experiments that the mass differences of the three oscillating active neutrinos are too small to be resolved by the present β-spectrometers. Nevertheless, these instruments established that the effective mass of the electron neutrino is smaller than 2 eV [8]. There are indications from a few neutrino oscillation experiments and cosmological observations that, in addition to the active neutrinos $\nu_e$, $\nu_\mu$ and $\nu_\tau$, there could exist sterile neutrinos, e.g. refs. [9], [10], [11]. As we describe in Section 3.7.2, the upcoming β-spectrometers could provide useful information about these hypothetical particles in the eV mass range and maybe also in the keV mass range.

Neutrino physics is an active research field with great consequences for particle physics, astrophysics and cosmology. During the last 14 years since the undoubted proof of the neutrino oscillations [12], 600 to 800 preprints involving the word "neutrino" in their title were deposited each year in the arXiv of the Cornell University library. Recent reviews of neutrino physics are available, both general (e.g. [13], [14], [15], [16], [17] [18]) as well as specialized[4]. Newer monographs (e.g. [19], [20], [21], [22], [23]) have also appeared that complement the earlier book [7] of Boehm and Vogel on massive neutrinos. Concise reviews have been regularly issued by the Particle Data Group [8]. Current problems in the theory of the neutrino masses were recently summarized in ref. [24].

---

[4] Examples are the recent reviews on neutrinos in astrophysics and cosmology [25], [26], [27], neutrino oscillations [11], [28], the 0νββ decay [29], [30], kinematic searches for the neutrino mass [31] and sterile neutrinos [10], [32].

In this review we first briefly compare the various methods used for the neutrino mass determination, present their current results and discuss signatures of active and sterile neutrinos in the β-ray spectra. There is no need to describe all previous β-spectra measurements since several outstanding reviews are available, e.g. [31], [33], [34], [35], [36]. Here we focus on electron spectroscopy aspects of the neutrino mass determination.

## 2. EXPERIMENTAL METHODS YIELDING INFORMATION ON THE NEUTRINO MASS

Some of the applied methods (often called kinematic or direct) rely only on the energy and momentum conservation in various weak decay processes. Other methods can be more sensitive but their results depend on assumptions of applied nuclear or cosmological models. All of these methods are needed since they are complementary and independent of each other. The requirement for mutual consistency in their results is thus a strong test of possible measurement errors as well as a test of theory.

**2.1 Neutrino oscillations**

The Standard Model of particle physics assumes that all matter in the universe is composed of leptons and quarks. With the exception of neutrinos, the masses of these basic fermions are known [8]. Although not required by any law of physics, the mass of neutrinos used to be assumed to equal zero. However, neutrino oscillation experiments revealed that it is not the case. According to present knowledge (see e.g. [19]) the neutrino $\nu_\alpha$, created with flavor α in a charged-current weak interaction process, is a quantum superposition of the neutrino mass eigenstates $\nu_i$

$$\nu_\alpha = \sum_i U_{\alpha i} \nu_i \qquad (1)$$

Here, the index α = e, μ, τ marks three known kinds of the active[5] flavor neutrinos $\nu_e$, $\nu_\mu$ and $\nu_\tau$. The index $i$ denotes the individual neutrino mass eigenstates with mass $m_i$. $U_{\alpha i}$ are the elements of the Pontecorvo-Maki-Nakagawa-Sakata neutrino mixing matrix. Their absolute values are known from neutrino oscillation experiments with rather good accuracy [8]. Oscillation experiments also yield the differences of squares of the neutrino masses, $\Delta m_{jk}^2 = m_j^2 - m_k^2$, or at least their absolute values. Since measured $\Delta m_{jk}^2$ do not equal zero a fact of topical importance follows: at least two neutrino mass eigenvalues are different from zero. However, at the time of writing the masses $m_i$ themselves remain unknown and the same holds for the mass ordering where all three options are likely: $m_1 < m_2 < m_3$ [the normal hierarchy (NH)], $m_3 < m_1 < m_2$ [the inverted hierarchy (IH)], or $m_1 \simeq m_2 \simeq m_3$ [the quasi-degenerate mass spectrum]. With one exception [37] that has not yet been confirmed by any other experiment, almost 70 years of effort by experimentalists has resulted in knowing only the lower and upper limits of the neutrino masses $m_1$, $m_2$ and $m_3$ or their combinations.

Oscillation experiments carried out with solar, atmospheric, accelerator and reactor neutrinos have proved to be an extremely powerful tool of neutrino physics (see e.g. [28]) since they have demonstrated that neutrinos are massive particles. From current values of the squared mass differences $\Delta m_{21}^2$ = (7.53 ± 0.18) ·10$^{-5}$ eV$^2$, $|\Delta m_{32}^2|$ = (2.44 ± 0.06) ·10$^{-3}$ eV$^2$ for NH and $|\Delta m_{32}^2|$ = (2.52 ± 0.07) ·10$^{-3}$ eV$^2$ for IH [8], it is possible to derive a lower limit on two of the mass eigenvalues. The limits (at the 90 % CL) are $m_2$ > 8 meV, $m_3$ > 49 meV for NH and $m_1$ > 48 meV, $m_2$ > 49 meV for IH. Of course the real masses $m_i$ can be much larger, i.e. it may hold that $m_1 \simeq m_2 \simeq m_3$. Possible values of the neutrino masses are depicted in Fig. (**1**).

---

[5] As described in Sect. 3.7, additional sterile neutrinos have been suggested to explain anomalous results of several oscillation experiments and cosmological observations.

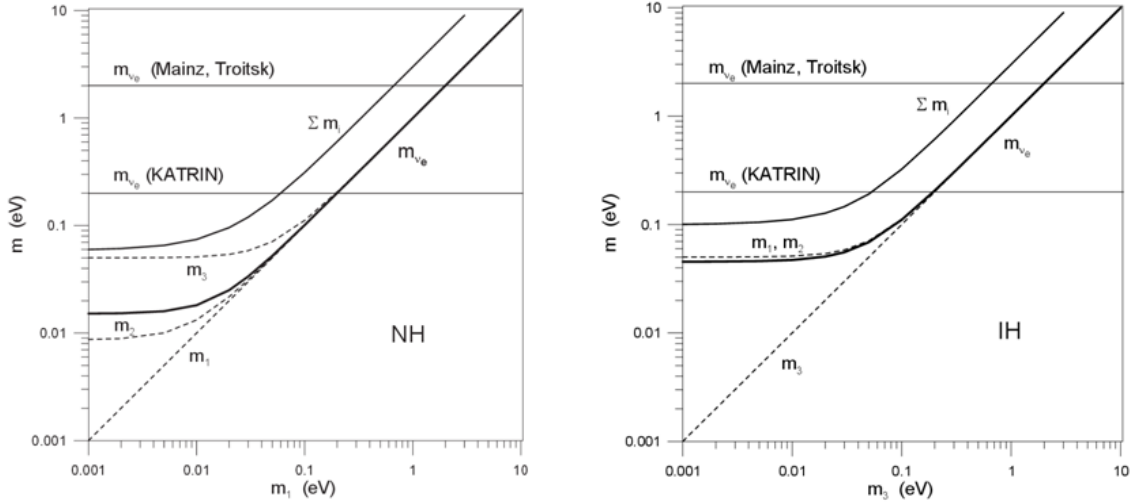

**Fig. (1).** Neutrino masses $m_i$ as a function of the lightest mass $m_1$ for the normal hierarchy NH (left panel) and $m_3$ for the inverted hierarchy IH (right panel). We have calculated the curves for the mean values $\Delta m_{21}^2 = 7.53 \cdot 10^{-5}$ eV², $|\Delta m_{31}^2| \approx |\Delta m_{32}^2| = 2.44 \cdot 10^{-3}$ eV² for NH and $2.52 \cdot 10^{-3}$ eV² for IH, and for the admixtures $|U_{e1}|^2 = 0.6803$, $|U_{e2}|^2 = 0.2354$ and $|U_{e3}|^2 = 0.0843$ derived from ref. [8]. Also shown is the effective neutrino mass $m_{\nu_e}$, the β-spectroscopic quantity defined by Eq. (5), as well as the sum of neutrino masses $\sum m_i$ derived from cosmological measurements. Also depicted is the upper limit of $m_{\nu_e}$ resulting from the latest measurements at Mainz [38] and Troitsk [39] and the expected limit from the upcoming KATRIN experiment [40].

## 2.2 Neutrinoless double beta decay

If the neutrino is a Majorana particle (i.e. neutrino ν and antineutrino $\overline{\nu}$ are identical particles) the 0νββ decay, e.g. $(A, Z) \rightarrow (A, Z+2) + 2e^-$, could develop into an important tool for determination of the effective Majorana neutrino mass, namely

$$m_{\beta\beta} = |\sum_k m_k \cdot |U_{ek}|^2 \cdot e^{i\varphi(k)}|. \qquad (2)$$

Here $\varphi(k)$ are the yet unknown Majorana phases, $0 \leq \varphi(k) \leq 2\pi$. Assuming that the 0νββ decay proceeds dominantly via the emission and subsequent absorption of a virtual neutrino, its half-life can be expressed as

$$\left(T_{1/2}^{0\nu\beta\beta}\right)^{-1} = G_{0\nu} \cdot |M_{0\nu}|^2 \cdot m_{\beta\beta}^2. \qquad (3)$$

The kinematic factor $G_{0\nu}$ is exactly calculable, in contrast to the value of the nuclear matrix element $M_{0\nu}$, which depends on the applied nuclear model. Moreover, it is necessary to take into account all virtual states of the intermediate nucleus $(A, Z+1)$, most of which are not known from experiment. A comparison of different evaluations indicates a spread by a factor of 2 to 3 in the calculated nuclear matrix elements [41].

The current state of the 0νββ decay theory was recently reviewed in ref. [42]. The authors argue that it would be possible to disentangle various mechanisms that may contribute to the neutrinoless double β-decay if this decay were examined in a sufficient number of isotopes. However, other investigators [43] recommend a more cautious interpretation of the experimental data of $T_{1/2}^{0\nu\beta\beta}$. According to ref. [44], the matrix elements from present calculations differ from one to another by up a factor of three. In addition to the possibility of experimental determination of the effective neutrino mass $m_{\beta\beta}$ according to Eq. (2), the 0νββ decay seems to be the only known way to examine a possible violation of the law of lepton number conservation. Long-standing searches for this extremely rare process are described in e.g. refs. [45], [46]. Requirements for current and future 0νββ experiments are discussed in [47], [48], while new expectations and uncertainties in 0νββ decay are reviewed [49] in the light of recent neutrino oscillation studies as well as cosmological observations. The latest results of the double beta decay searches along with the expected sensitivities of experiments in construction are reviewed in refs. [50], [51] and [52].

With one exception, all the searches until now have found only a lower limit of $T^{0\nu\beta\beta}_{1/2}$, i.e. the upper limit of $m_{\beta\beta}$. The only exception is the claim in ref. [53] for an observation of the $^{76}_{32}Ge \rightarrow ^{76}_{34}Se + 2e^-$ decay with a half-life $T^{0\nu\beta\beta}_{1/2} = (2.23^{+0.44}_{-0.31}) \cdot 10^{25}\ y$. Assuming dominance of the 0νββ-decay mode and choosing a particular value of the nuclear matrix element $M_{ov}$, from Eq. (3) the authors derived[6] the effective mass $m_{\beta\beta} = (0.32^{+0.03}_{-0.03})$ eV [37]. The GERDA experiment [56], aiming to check this claim, recently reported the results of its phase I: $T^{0\nu\beta\beta}_{1/2}(^{76}Ge) > 2.1 \cdot 10^{25}$ y at 90% CL [57]. An example of the GERDA spectrum is shown in Fig. (**2**). Results of the recent searches for the 0νββ decay are summarized in Tab. (**1**).

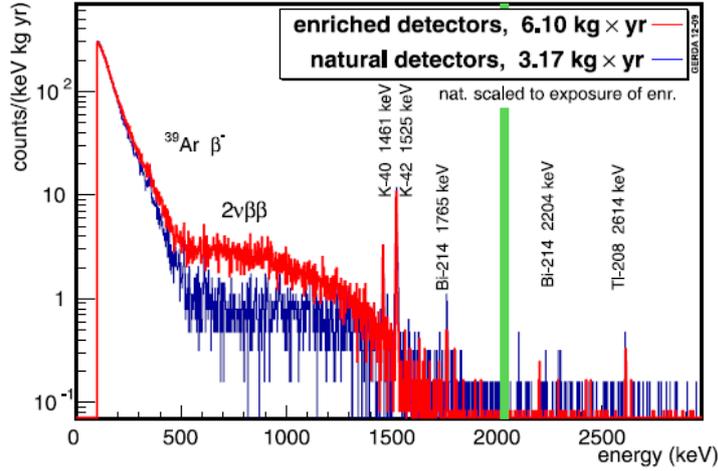

**Fig. (2).** Spectrum of electrons and photons recorded in the GERDA experiment [56] with germanium detectors enriched in $^{76}$Ge (red spectrum with higher intensity) and non-enriched Ge detectors (blue spectrum with lower intensity). The continuous spectra correspond to the sum of energy of two electrons released in the 2νββ decay of $^{76}$Ge with the half-life of $1.5 \cdot 10^{21}$ y. The low energy part of the spectrum is made mostly by β particles from the $^{39}$Ar decay and photon bremsstrahlung of these particles. Pure β emitters $^{39}$Ar ($T_{1/2}$ = 269 y, $Q_\beta$ = 565 keV) and $^{42}$Ar ($T_{1/2}$ = 33 y, $Q_\beta$ = 600 keV with daughter $^{42}$K) are produced via cosmogenic activation of atmospheric argon. The specific activity of $^{39}$Ar is 1 Bq kg$^{-1}$. Spectral lines with higher energy are due to γ quanta emitted in a decay of natural radioisotopes. Vertical band at $Q_{\beta\beta}$ = 2039 keV with the bandwidth of ±20 keV denotes the region where the line from the 0νββ decay is expected. The background in this region is about 0.02 counts keV$^{-1}\cdot$kg$^{-1}\cdot$y$^{-1}$.

**Table 1. Lower limits of the $T^{0\nu\beta\beta}_{1/2}$ half-lives determined at the 90% CL in recent experiments along with upper limits of the effective neutrino mass, $m_{\beta\beta}$, derived for various values of the nuclear matrix elements.**

| Experiment | Isotope | $T^{0\nu\beta\beta}_{1/2}$ (y) | $m_{\beta\beta}$ (eV) |
|---|---|---|---|
| GERDA [57] | $^{76}$Ge | $> 2.1 \cdot 10^{25}$ | $< (0.2 - 0.4)$ |
| NEMO-3 [58] | $^{100}$Mo | $> 1.1 \cdot 10^{24}$ | $< (0.3 - 0.8)$ |
| CUORICINO [59] | $^{130}$Te | $> 2.8 \cdot 10^{24}$ | $< (0.30 - 0.71)$ |
| EXO-200 [60] | $^{136}$Xe | $> 1.1 \cdot 10^{25}$ | $< (0.19 - 0.45)$ |
| KamLAND-Zen [61] | $^{136}$Xe | $> 1.9 \cdot 10^{25}$ | $< (0.12 - 0.25)$ *) |

*) Combined with the former result of the EXO-200 experiment [62].

---

[6] The interpretation of their spectra [53] has been subjected to criticism, see e.g. [54], [55].

## 2.3 Cosmological observations

Important, though strongly model-dependent, information about the neutrino mass states can be gained from cosmological data. Of particular relevance are the precision measurements of temperature fluctuations of the cosmic microwave background (CMB) and investigation of large scale structures (LSS) of the universe [19]. The extensive sets of measured data enable the estimation not only of the usual six parameters of the standard cosmological model, but also of the sum of the neutrino masses $\sum m_i$ and the effective number of the neutrino families $N_{eff}$. Details can be found, e.g. in reviews [63], [64] presented at the Neutrino 2014 conference. Theoretical aspects of interconnection between neutrino physics and dark matter issues have most recently been considered in ref. [65].

Neutrino oscillation experiments yielded a lower limit $\sum m_i \geq 0.05$ eV. The cosmological upper bounds of $\sum m_i$ depend on the types of measurements included and details of their analysis. For example, the newest results of the Planck collaboration range from $\sum m_i < 0.4$ eV up to $\sum m_i < 1.7$ eV at the 90% CL [66]. The often quoted estimate $\sum m_i < 0.6$ eV is considered to be a conservative one. On the other hand, Beutler *et al.* [67] derived from the current cosmological data the sum of the neutrino masses $\sum m_i = 0.36 \pm 0.10$ eV (68% CL). This estimate, together with the known elements of the neutrino mixing matrix [68], led to the values of the kinematically measured quantity $m_{\nu_e}$ (defined in Section 2.4) equal to 0.117 ± 0.031 eV for NH and 0.123 ± 0.032 eV for IH [67]. Most recently, Palanque-Delabrouille *et al.* analyzed the measured Lyα forest power spectra combined with the spectra of CMB and baryon acoustic oscillation data. The authors assumed validity of the ΛCDM cosmological model in the presence of massive neutrinos. Altogether, 7 cosmological, 6 astrophysical and 13 nuisance parameters were varied simultaneously. Both frequentist and Bayesian approaches were applied and the stability of the solution was examined. The analysis yielded the strongest cosmological constraint on the total neutrino mass, $\sum m_i < 0.14$ eV at the 95% CL [69]. Detailed discussion of neutrino constraints resulting from data on LSS and CMB can be found in ref. [70]. Constraints on the $\sum m_i$ expected from future observations of galaxy clustering and cosmic microwave background have recently been treated in ref. [71].

## 2.4 Kinematic methods

In principle, the kinematic methods could yield a $m_i$ value for each of the neutrino mass states assuming that this state is involved in the corresponding weak-interaction state with a measurable intensity, and assuming that the applied instrument has sufficient energy resolution to separate individual mass states. However, the current oscillation experiments show that the splitting of the mass states is far too small to be resolved by current spectrometers. In case of β- decay, various effective neutrino masses are presented in ref. [72], [73], [74]. In accord with the Particle Data Group [8] we utilize the following definition of the effective neutrino mass $m_{\nu_\alpha}$

$$m_{\nu_\alpha}^2 = \sum_i |U_{\alpha i}|^2 \cdot m_i^2 \qquad (4)$$

Ref. [8] gives the following limits for the electron-based, μ-based and τ-based neutrino masses[7], respectively

$$m_{\nu_e} = \sqrt{\sum_i |U_{ei}|^2 \cdot m_i^2} \; < 2 \text{ eV} \qquad (5)$$

$$m_{\nu_\mu} = \sqrt{\sum_i |U_{\mu i}|^2 \cdot m_i^2} \; < 190 \text{ keV} \qquad (6)$$

$$m_{\nu_\tau} = \sqrt{\sum_i |U_{\tau i}|^2 \cdot m_i^2} \; < 18.2 \text{ MeV}. \qquad (7)$$

The limit of $m_{\nu_e}$ in Eq. (5) originates from recent investigations of tritium β-spectra where the conservation of energy and momentum is applied [38], [39]. The limit in Eq. (6) is based on the precision measurement of the muon momentum in the two-body decay of pions at rest: $\pi^+ \rightarrow \mu^+ + \nu_\mu$ or $\pi^- \rightarrow \mu^- + \bar{\nu}_\mu$ [75]. The limit in Eq. (7) was derived from the kinematics of the $\tau^-$ decay [76]. The large differences among these limits reflect the absolute precision in momentum and energy determination achievable in the relevant experiments. Considering the relation among the total energy, momentum and the rest mass of a particle (in units $c = \hbar = 1$):

$$E_{tot}^2 = p^2 + m^2 \quad (8)$$

one can hope for a high sensitivity for the neutrino mass determination only when the neutrino is almost nonrelativistic; otherwise the experimental precision is lost in the $E_{tot}^2 - p^2$ difference, as underlined in [36].

---

[7] In accord with the CPT theorem it is assumed that the antineutrino mass is the same as the neutrino mass.

The mass of a particle can be determined via time-of-flight, which is also a kinematic method. However, for neutrinos with their small mass, extremely large distances between the source and the detector are necessary. Investigation of the neutrinos emitted by the supernova SN1987A yielded the upper limit $m_{\nu_e}$ < 5.7 eV at the 95 % CL [77]. It should be noted that this result depends on assumptions of the supernova models that are not yet capable of predicting the time-spread of the neutrino emission with sufficient precision. An extensive analysis of the SN1987A data has also been given in ref. [78].

**2.5 Comparison of experimental methods**

The relationship between the results of neutrino oscillation experiments, 0νββ-decay searches, cosmological observations and examination of β-spectrum shape has been explored by several investigators; see e.g. [41], [55], [79], [80]. The results of neutrino oscillation experiments enable not only the examination of the dependence of various neutrino quantities on the minimal neutrino mass $m_{min}$ as shown in Fig. (1) but also the exploration the relationship of these quantities. Newer graphical comparisons of $m_{\nu_e}$, $m_{\beta\beta}$, and $\sum m_i$ are depicted in Figs. (**3**) and (**4**).

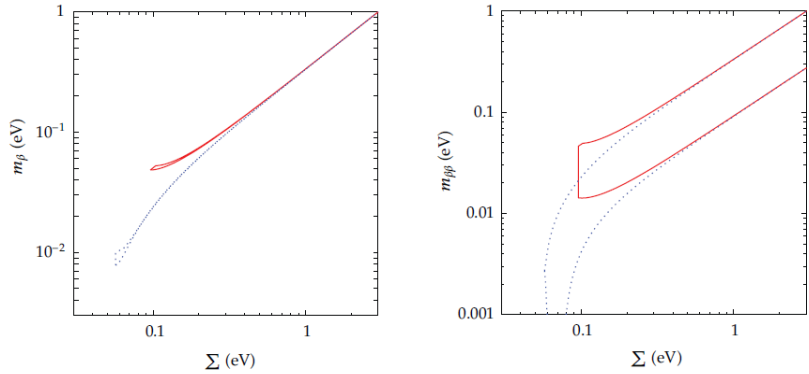

**Fig. (3).** Regions of the three observables sensitive to the absolute neutrino masses that are allowed by the neutrino oscillation experiments: $m_\beta \equiv m_{\nu_e}$ (Eq. 5), $m_{\beta\beta}$ (Eq. 2) and $\sum m_i$ for NH (blue dotted) and IH (red solid). Bands correspond to 3σ uncertainties of the oscillation data. Figure is reproduced from ref. [81].

Fig. (3) shows that the strongest cosmological constraint on total neutrino mass, $\sum m_i$ < 0.14 eV at the 95% CL [69], corresponds to the following limits on both $m_{\nu_e}$ and $m_{\beta\beta}$ of < 0.04 eV for NH and <0.06 eV for IH. According to ref. [41] and Fig. (**4**), the neutrino oscillation data indicate a lower limit of $m_{\beta\beta}$ for IH but no such limit for NH. The overlap of bands for NH and IH demonstrates that future 0νββ experiments could examine the neutrino mass ordering only if their sensitivity is better than 0.01 eV. In addition, the oscillation data allow the testing of whether the observed values of $m_{\beta\beta}$ and $\sum m_i$ are in agreement with the assumptions of the 3-neutrino framework and the dominance of the light neutrino exchange in the 0νββ-decay.

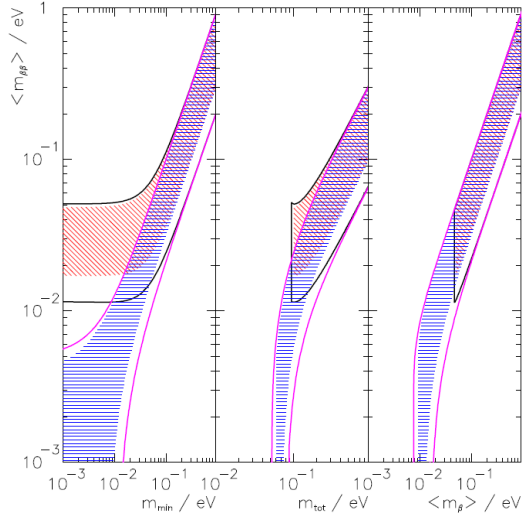

**Fig. (4).** Regions of the effective Majorana neutrino mass $m_{\beta\beta}$ allowed by the neutrino oscillation experiments, depending on the assumed values of the smallest neutrino mass $m_{min} = \min(m_i)$, $m_{tot} = \sum m_i$ and $<m_\beta> = m_{\nu_e}$ defined in Eq. 5. The blue horizontally-hatched band corresponds to NH, the slant red one to IH. The width of these bands reflects the uncertainty due to the unknown phases $\varphi(k)$ in Eq. 2. Wider bands restricted by solid lines (in red for NH and in black for IH) include the experimental uncertainties of the oscillation parameters. The systematic uncertainties of the nuclear matrix elements are not included. This figure is reproduced from ref. [41].

## 3. β-SPECTROSCOPIC DETERMINATION OF THE NEUTRINO MASS

### 3.1 Sensitivity of the neutrino mass to the β-spectrum shape

The first successful theory of the nuclear β-decay was developed by Fermi [2] who demonstrated that the shape of the uppermost part of the β-spectrum (close to the maximum energy of β-particles and thus to the minimum energy of neutrinos) is sensitive to the neutrino mass squared. Assuming only one neutrino mass state and no excitation of the daughter ion, the β-spectrum of allowed β-transitions ($\Delta I_{nucl} = 0, 1$ and $\Delta \pi_{nucl} = 0$) takes its simplest form[8] (see e.g. [31]):

$$\frac{dN}{dE} = A \cdot F(E, Z+1) \cdot p \cdot (E + m_e) \cdot \varepsilon \cdot \sqrt{\varepsilon^2 - m_\nu^2} \cdot \theta(\varepsilon - m_\nu) \tag{9}$$

Here $A$, $E$, $p$ and $m_e$ denote the spectrum amplitude and the electron kinetic energy, momentum and rest mass, respectively. $\varepsilon = E_0 - E$, where $E_0$ is the endpoint energy, i.e. the maximal electron kinetic energy for $m_\nu = 0$. The Fermi function $F(E, Z+1)$ takes into account the Coulomb interaction of the emitted β-particle with the daughter nucleus and its surrounding electrons. $\Theta$ is the Heaviside step function.

The Kurie plots in Fig. (**5**) show that the neutrino mass can be extracted from measured β-spectrum only if the spectrometer resolution $\Delta E_{instr}$ is comparable or better than $m_\nu$. In addition, the spectrometer must exhibit a high luminosity[9] and low background in order not to miss a weak effect at the upper end of the β-spectrum. Simultaneous fulfillment of these requirements was not yet achieved to the required level, despite more than half a century's effort. Thus all the measurements of the β-spectra up to now have yielded only an upper limit on $m_\nu$.

When the technique of β-ray spectroscopy was significantly improved by Bergkvist in 1972 [83], it became necessary to consider that a part of the β-decay energy may be spent for an excitation of the daughter ion. In this case the observed β-spectrum would be described as a superposition of partial β-spectra with endpoint energies $E_{0j} = E_0 - m_\nu - V_j$, where $V_j$ is the energy of the $j$-th excitation state populated with the probability $P_j$:

$$\frac{dN}{dE} = A \cdot F(E, Z+1) \cdot p \cdot (E + m_e) \cdot \sum_j P_j (\varepsilon - V_j) \cdot \sqrt{(\varepsilon - V_j)^2 - m_\nu^2} \cdot \theta(\varepsilon - V_j - m_\nu) \tag{10}$$

---

[8] The mathematical description of various theoretical correction terms is summarized in ref. [82].
[9] High luminosity $L = \Omega/4\pi \cdot S$, where $\Omega/4\pi$ is the fractional input solid angle and $S$ is the source area, is indispensable. The sources must be extremely thin to avoid deterioration of measured spectra by losses of electron energy within the source material.

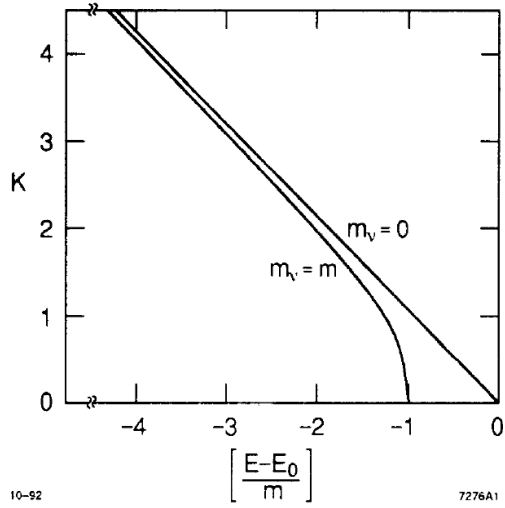

**Fig. (5).** Kurie plots for an allowed β-decay and two values of $m_\nu$ demonstrating narrowness of the endpoint region sensitive to the neutrino mass [35]. The graphs are obtained from Eq. (9) expressing the parameter $E - E_0$ in units of $m$ equaled to assumed nonzero value of $m_\nu$. Explicitly, $K \cong (E - E_0)/m \times [1 - m^2/(E - E_0)^2]^{1/4}$ for $m_\nu \neq 0$ and $K \cong (E - E_0)/m$ for $m_\nu = 0$.

In the case of β-decay of an isolated tritium atom, i.e. for a hydrogen-like $^3\text{He}^+$ ion, the quantities $V_j$ and $P_j$ can be precisely calculated (e.g. ref. [84]). The β-decay of a gaseous tritium molecules ($T_2$) is more complicated. Nevertheless, reliable calculations for the molecular ion $(^3\text{HeT})^+$ are available that take into account both electronic and rotational-vibrational molecular states, see Fig. (**6**). These predictions have not yet been experimentally verified but their precision and internal consistency indicates their applicability in the newest neutrino mass measurements. On the other hand, calculations of $V_j$ and $P_j$ for β-decay of atoms in many-electron systems of complex molecules or for β-emitters imbedded into solids may also suffer from insufficient knowledge of the composition of the radioactive samples. Very recently, a detailed assessment of molecular effects on the determination of $m_{\nu_e}$ from tritium β-spectra has become available [85].

Since the discovery of neutrino oscillations, it is necessary also to consider the individual neutrino mass states in the analysis of the measured β-spectra using the formula

$$\frac{dN}{dE} = A \cdot F(E, Z+1) \cdot p \cdot (E + m_e) \cdot \sum_j P_j \cdot (\varepsilon - V_j - m_i) \cdot \sum_i |U_{ei}|^2 \cdot \sqrt{(\varepsilon - V_j)^2 - m_i^2} \cdot \theta(\varepsilon - V_j - m_i)$$

(11)

This is exemplified in Fig. (**7**) for tritium β-decay, assuming the normal neutrino mass hierarchy with $m_1 = 200$ meV. Since the best instrumental resolution is still much worse than the differences among the individual neutrino mass states, the measured spectra can still be analyzed with the help of Eq. (10), where the variable $m_\nu$ should be interpreted as the effective electron neutrino mass $m_{\nu_e}$, defined in Eq. (5).

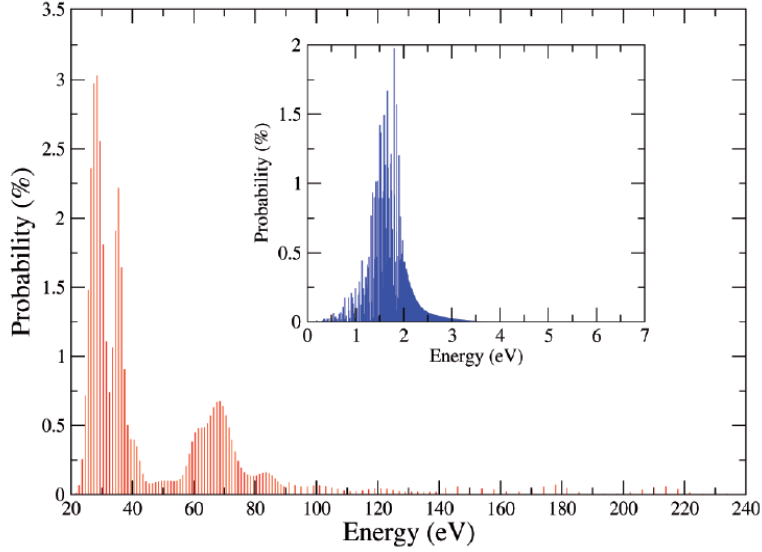

**Fig. (6).** Probability distribution for excitations of the ($^3$HeT)$^+$ ion populated the in β-decay of gaseous T$_2$ [86]. The figure combines results from the calculations of excitation to the electronic bound states [87] and to the electronic continuum [88] of ($^3$HeT)$^+$. The electronic ground state is populated in 57 % of the decays, 29 % of the decays end in the excited electronic states and remaining 14 % lead to the electronic continuum. The rotational-vibrational excitations of the molecular ion in its ground electronic state are shown in the insert. The mean energy of these excitations is 1.7 eV with the energy spread (FWHM) of about 0.4 eV. The electronic excitations of the ($^3$HeT)$^+$ ion starting at about 20 eV are shown in the lower part of the figure.

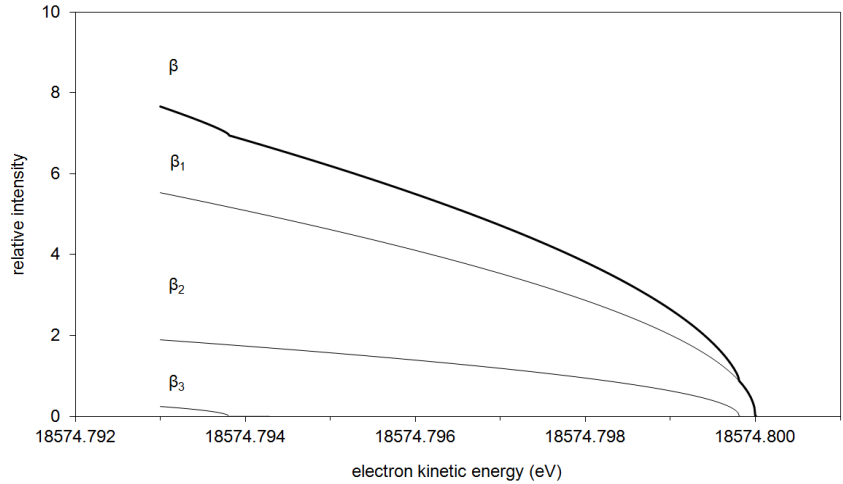

**Fig. (7).** The uppermost part of the tritium β spectrum, wide only 7 meV, was calculated for the normal hierarchy of the neutrino mass states ($m_1 < m_2 < m_3$) assuming the endpoint energy $E_0 = 18.575$ keV and $m_1 = 200$ meV. The same mass square differences $\Delta m_{ij}^2$ and the admixtures $|U_{ei}|^2$ as in Fig. (**1**) were assumed. Under these assumptions $m_2 = 200.19$ meV, $m_3 = 206.19$ meV, and the relative intensities of partial spectra β$_1$, β$_2$ and β$_3$ are $|U_{e1}|^2 = 0.6803$, $|U_{e2}|^2 = 0.2354$ and $|U_{e3}|^2 = 0.0843$, respectively [8]. The experimentally observable β-spectrum is composed of the three components. Taken with sufficiently high instrumental resolution, the spectrum will exhibit kinks at energies $E_0 - m_i$ ($i = 2, 3$). In our example, the average electron neutrino mass $m_{\nu_e}$ defined by Eq. (5) equals 200.57 meV. We have assumed for simplicity that all the $^3$H $\rightarrow$ $^3$He + e$^-$ + $\bar{\nu}_e$ decays proceed to the ground state of the daughter ion.

In order to compare a measured β-spectrum $s_{exp}(E)$ with theoretical predictions for various neutrino masses $s_{th}(E', m_{\nu_e})$, it is necessary to know the spectrometer resolution function $R(E,E')$. This function is the response of the spectrometer to a monoenergetic signal of unit intensity

$$s_{exp}(E) = \int R(E, E') \cdot s_{th}(E', m_{\nu_e}) \, dE' \qquad (12)$$

where integration covers the region with $s_{th}(E', m_{\nu_e}) \neq 0$. Precise determination of $R(E,E')$ for a real spectrometer turns out to be an non-easy task of great importance for reliable determination of $m_{\nu_e}$.

Analyses of the measured β-spectra are usually performed by the least-squares method with at least four fitted parameters: $m_{\nu_e}^2$, $E_0$, spectrum amplitude $A$ and background $b$. When fitting only the uppermost part of the β-spectrum, the uncertainty in the absolute energy scale of the spectrometer does not introduce a recognizable systematic error in the determination of $m_{\nu_e}$, see e.g. ref. [89]. The use of a fixed value of $E_0$, derived from precision measurement of the atomic masses of ³H and ³He, would greatly facilitate the task, since the $m_{\nu_e}^2$ and $E_0$ variables are strongly correlated. However, in this case the precision of both $E_0$ and the absolute energy calibration of the β-ray spectrometer have to reach a level of a few-meV [90], far from the present state of the art. Still, the comparison of fitted and independently measured $E_0$ values proves to be an important check of the correctness of the evaluation of the β-spectrum. The newest measurement of the cyclotron frequency ratios of ³He⁺ to HD⁺ and T⁺ to HD⁺ in the Penning ion trap have yielded a $Q$ value for tritium β-decay equal to 18 592.01 ± 0.07 eV [91], which will enable a sensitive test of the fitted value of $E_0$.

### 3.2 Early experiments: from $m_{\nu_e}$ < 5 keV to $m_{\nu_e}$ < 35 eV

When Pauli proposed the existence of neutrino in 1930 he estimated its mass "to be comparable with that of the electron, surely less than 1 % of the proton mass" [1]. Fermi analyzed the contemporary experimental β-spectra and concluded that the neutrino mass is much smaller than the electron mass, and most likely equal to zero [2].

The first upper limit on the neutrino mass, derived from a precision measurement of the shape of the β-spectrum, was obtained in 1948 by Cook *et al.* [92]. The authors investigated the β-spectrum of ³⁵S ($Q_\beta$ = 167.18 keV, $T_{1/2}$ = 87.5 d) using a magnetic spectrometer shown schematically in Fig. (**8**). From Kurie plots depicted in Fig. (**9**) the authors concluded that $m_{\nu_e}$ < 5 keV, i.e. less than one percent of the electron rest mass.

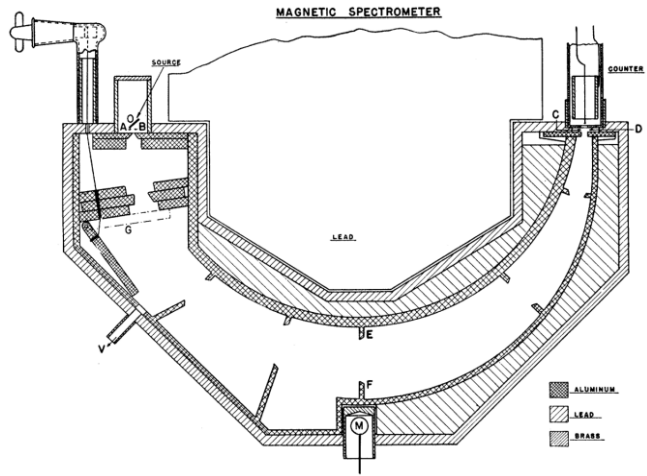

**Fig. (8)**. Magnetic spectrometer [93] with an electron orbit of 40 cm. The iron pole pieces produced the radially inhomogeneous magnetic field that yielded better focus at 180° than a homogenous field. The set of beam-defining slits made of aluminum minimized electron scattering. The instrument was adjusted for the relative momentum resolution[10] of 0.5 % and solid angle of about 0.1 % of 4π. The size of radioactive sources was (0.4 × 2.5) cm². The energy scale was calibrated with photoelectrons ejected from the K-shell of a lead convertor by the 511 keV annihilation radiation.

This upper limit of $m_{\nu_e}$ was already improved by 1949 owing to the development of proportional counters [94], [95]. These instruments enabled the measurement of the β-spectrum of gaseous tritium with extremely large geometrical transmission corresponding to the almost full solid angle of 4π. The low endpoint energy of tritium[11], $E_0 = 18.6$ keV, was advantageous, since the relative intensity of β-particles in the interval $\Delta E$ close to the endpoint is proportional to $(\Delta E/E_0)^3$. The resolution $\Delta E_{\text{instr}}$ of the applied counter [95] was 1.9 keV for the 17.4 keV calibration K$_\alpha$ x-rays of molybdenum. The results were $m_{\nu_e} < 0.5$ keV [96] and $m_{\nu_e} < 1$ keV [97].

---
[10] This momentum resolution $(\Delta p/p)_{instr}$ corresponds to the instrumental energy width $\Delta E_{instr}$ (FWHM) of 1.5 keV for $E$ = 167 keV.

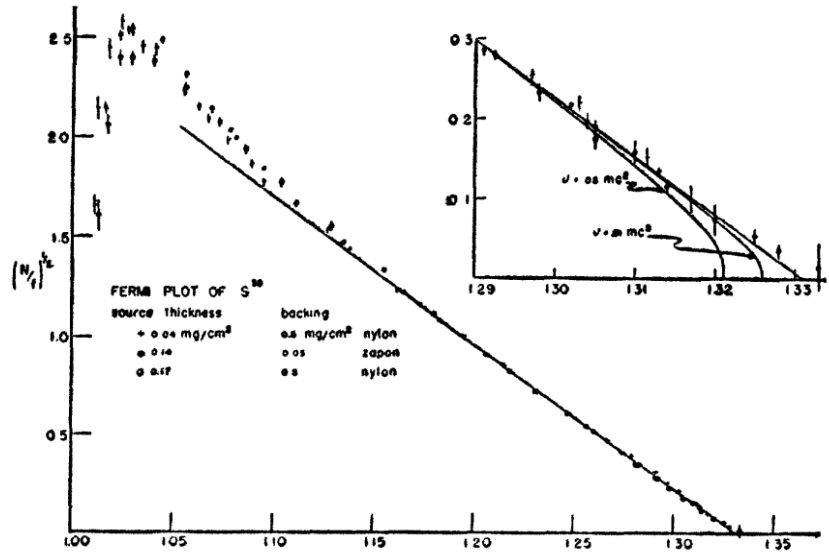

**Fig. (9)**. The Kurie plot of the β-spectrum of $^{35}$S measured by Cook *et al*. [92]. The plots shown correspond to the $^{35}$S sources of estimated thickness of 0.04, 0.14 and 0.17 mg cm$^{-2}$ that were prepared on 0.5 mg cm$^{-2}$ Nylon and 0.03 mg cm$^{-2}$ Zapon backings. For electron energies above 90 keV, the Kurie plots are all straight lines as expected for the $^{35}$S(3/2$^+$) → $^{35}$Cl(3/2$^+$) β-transition. According to the authors the deviations for energies below 60 keV are influenced by source thickness and backing. The enlarged section of the Kurie plot close to the β-spectrum endpoint shows the theoretical Fermi curves for an allowed β-transition calculated for assumed neutrino masses of 0, 5 and 10 keV.

Langer and Moffat [98] investigated the β-spectrum of tritium using the same magnetic spectrometer as was used in the previous studies of $^{35}$S [92]. The sources were prepared by thermal evaporation in vacuum of tritiated succinic acid (HOOC·CH$_2$CH$_2$·COOH) with specific activity of 33 MBq μg$^{-1}$ (i.e. about 10 % of protium atoms were substituted by tritium ones). This compound was quite stable at room temperature and sublimated at 150°C thus avoiding any thermal damage of a 4 μg cm$^{-2}$ thick organic support during evaporation. Autoradiography revealed a complete uniformity of the invisible deposit with an estimated surface density of about 0.5 μg cm$^{-2}$. Two methods were applied to prevent electrostatic charging of the source. First, a 2 μg cm$^{-2}$ thick and 1 cm-wide copper band was evaporated onto the reverse side of the insulating organic support. The band resistivity of 10$^8$ Ω was sufficient to insure good grounding. Second, an oxide-coated cathode filament was installed just below the source where electrons emitted from the heated filament would neutralize any positive charge created on the source due to the emission of β-particles. Both methods applied separately or together yielded the same endpoint energy $E_0$, while the one using an ungrounded source was lower by 460 eV due to the positive charging of the source. The analyses of measured β-spectra yielded an upper limit $m_{\nu_e} < 250$ eV [98].

In order to avoid difficulties with the detection of low-energy electrons, Hamilton *et al*. [99] investigated the tritium β-spectrum with a spherical electrostatic analyzer [100]. Instead of measuring counting rates, the authors recorded the current of β-particles that overcame a variable retarding potential and reached a spherical collector. From the integral spectrum measured with a rather thick tritium source (100 μg cm$^{-2}$) the authors derived an upper limit $m_{\nu_e}$ < 500 eV at 80 % CL. Their endpoint energy of $E_0$ = 17.6 ± 0.4 keV is 2.4 σ below its current value.

---

[11] Tritium has the second smallest endpoint energy of all known β-emitters. The super-allowed character of its β-decay leads to an acceptable half-life of 12.3 y. In suitable chemical forms, the simple electronic structure of tritium facilitates interpretation of the measured β-spectra. These are reasons why tritium was utilized in the majority of subsequent searches of $m_{\nu_e}$. $^{187}$Re with the lowest $E_0$ of 2.5 keV could not be examined via the usual β-spectroscopic methods due to its half-life of 4.3 · 10$^{10}$ y. However, the low-temperature micro calorimeters offer an interesting possibility for its investigation, see Sect. 3.5.

Salgo and Staub [101] examined the tritium β-spectrum with a retarding-potential spectrometer in a planar arrangement. The source was $T_2O$ ice evaporated on a gold plated holder of 5.5 cm radius that was kept at the temperature of liquid nitrogen. The $T_2O$ deposit of 39 μg mass and activity of 3.9 GBq was not homogenously distributed so that the surface density of about 20 μg cm$^{-2}$ estimated from interference fringes in the source center was 50 times higher than average density. Great attention was paid to secure a linear potential drop between the source and the planar retarding grid and to avoid high potential gradients near the source that could cause a field emission of electrons. Nevertheless, the background determined in the energy region above 18.75 keV with a statistical uncertainty of ± 0.056%, strongly dominated the effect. The results are $m_{\nu_e}$ < 200 eV at 80% CL and $m_{\nu_e}$ <320 eV at 90% CL.

Daris and St.-Pierre [102] investigated the tritium β-spectrum with an iron-core magnetic spectrometer adjusted for the resolution $\Delta p/p$ = 0.25 % (i.e. $\Delta E_{instr}$ = 90 eV at 18.6 keV) and fractional solid angle $\Omega/4\pi$ = 7.5 ·10$^{-4}$. Sources were made by absorption of tritium into an aluminum foil in a gaseous discharge of which the surface had been previously electropolished and de-oxidized. The sources, of dimensions 1mm × 54 mm, prepared at tritium pressure of 1.5 Torr and a discharge voltage of 600–700 V, had an activity of 0.7−1.2 MBq. Measurements of activity after successive electro peelings revealed that 90 % of the tritium ions were contained in a thickness of less than 7.4 nm (2 μg cm$^{-2}$) of aluminum. The energy loss of 18 keV electrons within the Al foil was estimated to be 20 eV. The instrumental resolution function was derived from ThC calibration lines. The resulting endpoint energy of 18.570 ± 0.075 keV agrees with present determinations. From four combined β-spectra, the authors concluded that $m_{\nu_e} = 0^{+75}_{-0}$ eV, where the error is the standard deviation σ. This means $m_{\nu_e}$ < 120 eV at the 90 % CL.

Bergkvist [103], [104] succeeded in increasing the luminosity of the π√2 magnetic spectrometer by two orders of magnitude without deterioration of the instrumental resolution. He developed an extended radioactive source, of which the parts were put on appropriate potentials in order to eliminate influence of the source width on the spectrometer resolution. The aberration defects were further reduced by an electrostatic corrector. For his study of $^3$H β-spectrum, Bergkvist [83] prepared the source, of (20 ×10) cm$^2$ area, by implantation of tritium ions of kinetic energy of 400–800 eV into an aluminium foil. The source also contained a set of small $^{170}$Tm sources of monoenergetic conversion electrons. These electrons of 22.9 keV kinetic energy, emitted from the K shell of daughter ytterbium atoms with natural widths $\Gamma$ =32 eV, enabled the determination of the instrumental resolution for the extended source to be $\Delta E_{instr}$ = 40 eV at 18.6 keV. In addition, the conversion electrons allowed the inspection of the stability of tritium source with respect to contamination over-layers that might develop during long runs.

Depth distribution of the tritium activity in the aluminum source backing was examined with the peeling technique. 93 % of activity was found in the 3 μg cm$^{-2}$ thick natural oxide layer while only 2.5 % of tritium ions penetrated deeper than 6.4 μg cm$^{-2}$. The overall experimental resolution, including energy losses within the tritium source, was estimated to be $\Delta E_{exp}$ = 55 eV at 18.6 keV. In spite of numerous slits in the spectrometer chamber, scattered β-particles produced about half of a high and energy dependent background above $E_0$ (see Fig. (**10**)). The second half was an intrinsic background of the large size detector needed for the extended source. The uppermost part of the measured tritium β-spectrum is shown in Fig. (**10**).

Bergkvist was the first who took into account that only 70 % of the tritium decays go to the 1s ground state of the $^3$He$^+$ daughter ion, while the remaining decays populate higher states, mainly the 2s state (see also [84]). This feature was approximated by means of two lines of the same width, shifted by 43 eV to each other. The width of the overall resolution function, utilized in the analysis of measured tritium β-spectrum, thus increased to 70 eV. This function was assumed to incorporate all the atomic and experimental effects, including energy losses in the source. The analysis of the measured tritium β-spectrum yielded $m_{\nu_e}$< 55 eV at the 90 % CL [83] and $E$ = (18610 ±16) eV [105].

Röde and Daniel [106], [107] investigated the tritium β-spectrum with the (π/2)√13 iron-free magnetic spectrometer [108] shown in Fig. (**11**).The instrument was adjusted for $\Delta p/p$ = 7.5 ·10$^{-4}$ ($\Delta E_{instr}$ = 28 eV at 18.6 keV) at a fractional solid angle of $\Omega/4\pi$ = 1.5 ·10$^{-2}$. A special anticoincidence proportional counter enabled the reduction of the background without the source inside the spectrometer to about 0.03 s$^{-1}$ for energies below 20 keV, regardless of the relatively large size of the detector.

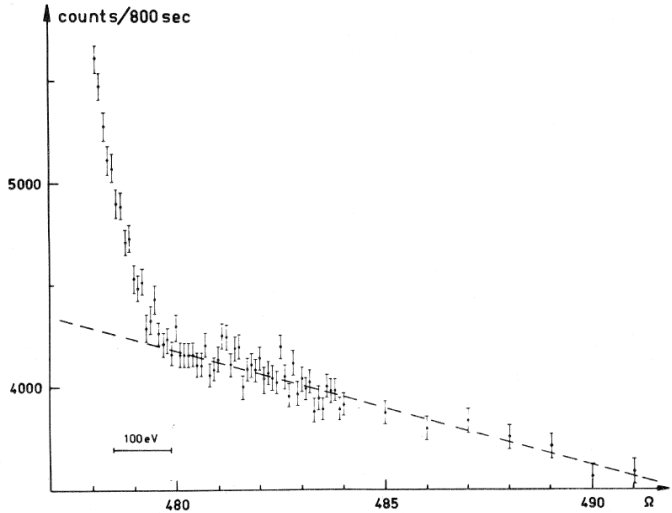

**Fig. (10).** The endpoint region of the tritium β-spectrum measured with the extended source in the Bergkvist experiment [83]. The quantity on the x-coordinate is proportional to the β-particle momentum. The spectrum above $E_0$ was checked for slight curvature by recording the background in the range 480 – 510 Ω. This test allowed the use of a sloping straight line for background description, as shown.

In order to increase the luminosity of their instrument the authors used a technique of an extended source [103]. Instead of a set of narrow strips at different potentials they applied an electrostatic correction by a continuous potential drop. For this purpose the 10 μg cm$^{-2}$ nitrocellulose backing of the radioactive layer was covered by 6 μg cm$^{-2}$ silver layer with a resistance of $10^9$ Ω between the adjacent potential defining wires with 1.2 mm spacing. A thin tritium film, prepared by spreading out a solution of tritiated polystyrene in benzene on water, was deposited on the silver side of the backing. Four tritium sources with dimensions (22 × 23) mm$^2$ and surface density of 4 – 6 μg cm$^{-2}$ were prepared in this way. Measurements of monoenergetic electrons emitted from a similarly extended $^{216}$Po source proved that the instrumental resolution achieved with these extended sources was the same as that for an equipotential source of (0.5 ×20) mm$^2$. The quality of one of the sources is demonstrated by a completely straight Kurie plot of the last 3 keV region of the β spectrum as shown in Fig. (12). The derived endpoint energy $E_0$ = (18.649 ± 0.074) keV agrees within 1σ with its current value and the upper limit $m_{\nu_e}$ < 86 eV was determined at the 90 % C. L. [106].

Tretyakov [109] developed a toroidal iron-free magnetic spectrometer which proved extremely useful in searches for $m_{\nu_e}$ in tritium β-spectra[12]. A magnetic field of the instrument was produced by 72 rectangular coils placed symmetrically around the spectrometer axis, which contained a radioactive source and the detector itself. This arrangement enabled the analysis of electrons emitted in the angular intervals of ±7.5° and twice 120° with respect to the source normal. To increase the dispersion of the device, electrons were led through the analyzing field four times, turning their direction by 180° in each path. This magnetic deflection by 720° also almost eliminated background caused by scattered electrons. In measurements of tritium β-spectra, Tretyakov *et al.* [113] utilized an extended non-equipotential source of the Bergkvist type [103] situated in the spectrometer focal plane. The composed source consisted of 18 parts, each of the size (2.6 × 18) mm$^2$. The parts were arranged in 9 pairs emitting electrons in opposite directions, thus enabling more efficient use of the spectrometer azimuthal angle.

The luminosity $L = S \cdot \Omega/4\pi$ = 0.07 cm$^2$ and resolution $\Delta p/p$ = 1.2 ·10$^{-3}$ were measured for a similar composed source emitting calibration conversion electrons from the $^{169}$Yb decay. The tritium sources were prepared by vacuum evaporation of tritiated valine, an amino acid (CH$_3$)$_2$CH·CH(NH$_2$)·COOH (where two of the protium atoms were substituted by tritium ones), at ~300°C on Al backing. The surface density of the source was estimated to be 2 μg cm$^{-2}$ of valine. Due to the small size of the low-pressure Geiger-Müller counter and the absence of scattered electrons in the detector region, the background above the tritium endpoint was 50 times smaller than that of the Bergkvist measurement shown in Fig. (**10**). In addition, the background did not depend on the electron energy and was the same regardless of the presence or absence of the tritium source in the spectrometer. Nevertheless, tritium itself slowly contaminated the spectrometer vacuum chamber, and the background after one month of measurements was four times higher than at the beginning.

---

[12] In addition to the first β-ray spectrometer of this type built in the Institute of Theoretical and Experimental Physics in Moscow, similar instruments were also constructed at the Los Alamos National Laboratory [110], at the Physics Institute of the University of Zurich [111] (Fig. (**15**)), and at the Lawrence Livermore National Laboratory [112].

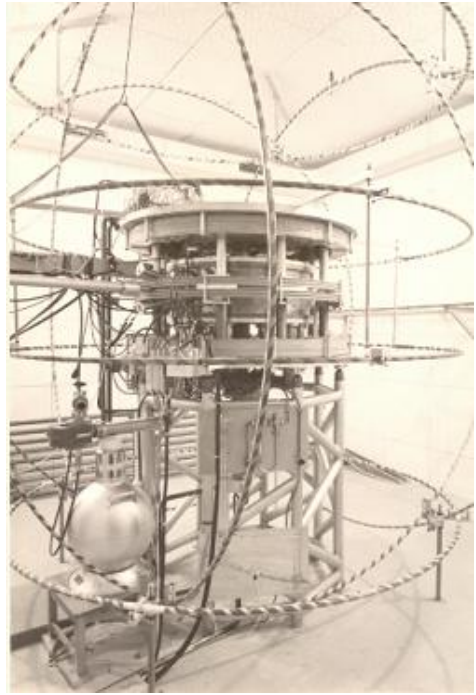

**Fig. (11).** An iron-free magnetic β-ray spectrometer with radial focusing and axial defocusing at the curved exit slit [108]. This focusing principle offers better combination of resolution and transmission than conventional π √2 spectrometers at the price of a larger detector and focusing angle of (π/2)·√13 = 324.5°. The radius of a central orbit is 30 cm. Note the coils compensating external magnetic fields in three perpendicular directions.

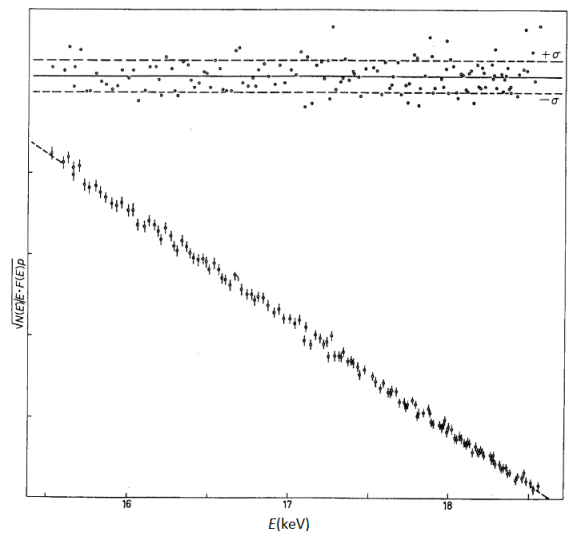

**Fig. (12).** The Kurie plot of the tritium β-spectrum measured by [106]. The 3 keV wide endpoint region is shown for a source with a thickness of the active layer of (4.5 ± 1.5) μg cm$^{-2}$. Quality of the fit is demonstrated by a spectrum of residuals in the upper part of the figure.

The least-squares fits of four series of the measurement of the β-spectrum based only on the statistical uncertainties of the counting rates yielded a normalized $\chi^2$ = 1.49 which was improbably large for 191 degrees of freedom. From the scattering of the counting rates in the overlapping regions of individual series, the authors estimated a total measurement uncertainty to be $\sigma_{tot} = 1.2\ \sigma_{stat}$. Two additional uncertainties were taken into account: ±15% for the uncertainty of the population of the effective excited states of the $^3$He$^+$ ion and 20 % due to a possible inhomogeneity of the source thickness. The Kurie plot of the recorded part of the tritium β-spectrum is reproduced in Fig. (**13**). The final result was $m_{\nu_e} <$ 35 eV at the 90 % CL [113]. The least-squares analysis of measured spectra yielded $E_0$ = (18574.8 ± 1.3) eV while caution consideration of systematic uncertainties led to $E_0$ = (18575 ± 13) eV. Both determinations agree perfectly with the current value of $E_0$.

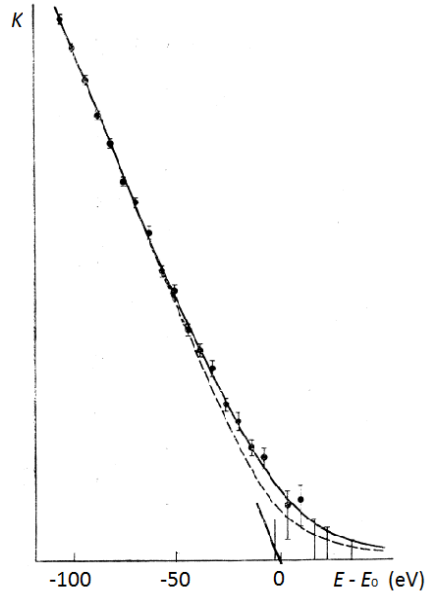

**Fig. (13).** Kurie plot of the endpoint region of the tritium β-spectrum measured with a toroidal magnetic spectrometer by Tretyakov *et al* [113]. Solid and dashed lines correspond to $m_{\nu_e}$ = 0 and 40 eV, respectively.

In order to reduce problems with electron scattering in radioactive sources and spectrometers as well as uncertainties connected with final state interactions, Simpson [114] investigated the β-spectrum of tritium implanted into a Si(Li) x-ray detector. The detector of 80 mm$^2$ area and 5 mm thickness had resolution of 215 eV at 5.9 keV. A diffuse beam of tritium ions of energy between 8 and 9.1 MeV penetrated through 25 μm thick Be window into 0.2 mm of Si(Li) crystal. This penetration depth far exceeded the 2.5 μm range of the 18.6 keV β-particles and 14 μm range of 1.7 keV Si x-rays. During implantation the detector was kept at liquid-nitrogen temperature. The detector resolution deteriorated at the end of implantation but returned to its original value after annealing. The atomic density of implanted tritium was estimated to be about 1.5·10$^{13}$ cm$^{-3}$. A chopper emitting 8.05 and 17.48 keV $K_{\alpha1}$-X rays of Cu and Mo respectively, was installed in front of the Si(Li) detector to enable energy calibration and stabilization during long tritium β-spectra runs. The chopper wheel with two slots allowing penetration of x-rays to the detector rotated with a period of a few seconds. In this way, two spectra were recorded with the same electronics setting almost simultaneously: a spectrum of β-rays plus x-rays and the other of β-rays only. The measured tritium β-spectra did not reveal any deviations from the theoretical shape expected for zero neutrino mass and the upper limit of $m_{\nu_e} <$ 65 eV was derived with 95% confidence.

**3.3 Claim for the 30 eV-neutrino and its disproval**

Physicists at the ITEP in Moscow continued to study tritium β-spectra using the same source material and the same toroidal magnetic spectrometer as in their previous study [113]. The application of a proportional counter decreased by a factor of 10 the background caused by a steadily growing contamination of the spectrometer with tritium. Later, a three-channel proportional chamber was installed that increased the rate of data accumulation. Analysis of the data collected during five years (including those from ref. [113]) led to an astonishing claim that $14 \leq m_{\nu_e} \leq 46$ eV at the 99 % CL [115][13]. This was the first claim for the non-zero neutrino mass. In the following seven years, the ITEP physicists improved their apparatus as well as their methods of β-spectra measurement and evaluation. A proportional chamber with six channels was installed as a detector. A non-equipotential source with continuous correction potential was built improving the instrumental resolution to $\Delta E_{instr} = 20$ eV at 18.6 keV. To achieve the continuous potential distribution the source backing was made of a weakly conducting glass with specific resistivity of $10^7$ Ω cm$^{-1}$. The β-spectrum was scanned by electrostatic field variation while the magnetic field was kept constant. In this way, an uncertainty in the dependency of the detector efficiency on the electron energy was eliminated. The magnetic field was adjusted to focus electrons of energy of 21.2 keV, exceeding the tritium endpoint energy. Nevertheless, the final result, published in 1987, remained substantially unchanged: $m_{\nu_e} = 30.3^{+2}_{-8}$ eV [117]. In the same work, "the model independent mass interval", $17 < m_{\nu_e} < 40$ eV was derived from the mass difference of the doublet $^3$H – $^3$He. However, the utilized mass difference [118] is in disagreement with more recent values [119], [91].

---
[13] Details of the work appeared later in the report [116].

The ITEP claim motivated several groups to develop new β-spectrometers and new source-preparation techniques. Extensive discussion of these experiments, including the now disproved ITEP result, can be found in the reviews [7], [33], [34], [35], [120], [121]. In the following, we mention a few details of possible interest in future β-spectroscopic searches for massive neutrinos.

In order to account properly for the resolution function of their magnetic spectrometer as well as for the electron-energy losses and back-scattering effect in their tritium source, Kawakami *et al.* [122] utilized the same chemical compound for the $^3$H source as well as for the $^{109}$Cd reference sources: an arachidic acid ($C_{20}H_{40}O_2$) in the form of its Cd salt. The former was labeled with tritium while the letter was labeled with $^{109}$Cd emitting monoenergetic KL$_2$L$_3$ Auger electrons of (18511.7 ± 1.3) eV energy. The sources of dimensions (20 × 6) cm$^2$ were deposited in two monomolecular layers of total thickness of 5 nm (0.6 μg cm$^{-2}$) on high-resistivity Ru$_2$O$_3$-coated aluminum plates. Optical aberrations due to the source size were compensated by suitable voltages on the source plate. A proportional chamber with resistive wires had position resolution of 0.5 mm for 18 keV electrons. The endpoint region of the tritium β-spectrum and the KL$_2$L$_3$ Auger line were recorded with the overall resolution of 30 eV. The overall response function of the whole setup $R_T$ shown in Fig. (**14**) was derived by deconvoluting a natural width ($\Gamma$) and the shake up/off effect (SUO) from measured Auger line:

$$R_T = (Auger) \otimes \Gamma^{-1} \otimes \text{SUO}^{-1}. \tag{13}$$

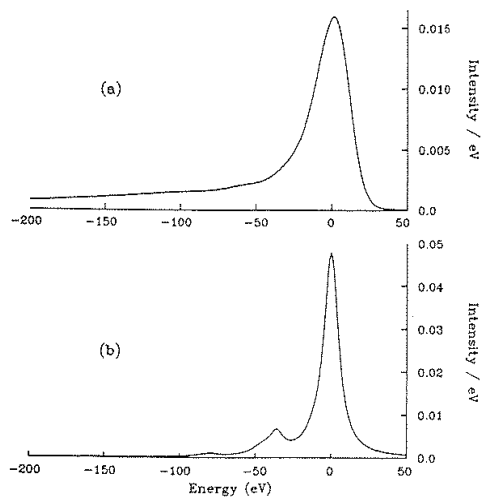

**Fig. (14)**. (a) The overall response function of the experiment [122] obtained from the measured KL$_2$L$_3$ Auger line by means of Eq. (13). The low-energy tail extends up to −1000 eV. The total intensity is normalized to one. (b) The calculated shake up/off spectrum convoluted with a Lorentzian corresponding to the natural line width $\Gamma = 11.43$ eV. The peak at the energy loss of 0 eV corresponds to the no-energy-loss component whose probability is 80%.

A major improvement in precision β-ray spectroscopy was achieved by Robertson *et al.* [110] at the Los Alamos National Laboratory, who applied a gaseous source with circulating molecular tritium. The main advantage was that the theoretical spectrum of final states of the (T$^3$He)$^+$ molecular ion is known more reliably than that of complex solid sources. Additionally, the energy losses of β-particles within tritium gas could be determined more precisely than in the case of solid sources. A solenoidal magnetic field guided to the spectrometer only those β-particles that did not backscatter on inner walls of the source vessel. Since no separation window was allowed between the source and the spectrometer, a steady flow of tritium into the source was needed, followed by an intensive differential pumping to avoid spectrometer contamination. No increase of background due to escaped tritium was observed and the background rate of ~0.005 $s^{-1}$ was caused mainly by cosmic rays. A toroidal magnetic spectrometer of the Tretyakov type [109] with focal length of 5 m, was adjusted to a fixed energy above the tritium endpoint and variable negative bias voltages were applied to the source tube to scan the spectra. Electrons were detected by an array of Si micro strip detectors, allowing the recording of 12 spectra simultaneously. Observed counting rates were renormalized for pressure variations in the source, the activity of which was monitored with another Si detector. Instrumental resolution was measured by circulating gaseous $^{83m}$Kr through the source and examining the shape of the K-conversion line with 17.8 keV. The natural width of this line, as well as shake-up and shake-off satellites, were carefully taken into account. The authors [110] also quantified 10 various contributions of uncertainties in the neutrino mass determination.

Holzschuh *et al*. [123] investigated the tritium β-spectrum with an iron-free toroidal magnetic spectrometer [111] that is shown in Fig. (**15**). The cylindrical source consisting of 10 disks (at appropriate potentials to compensate for the axial source extension) had a total active surface of 157 $cm^2$. Spectra were taken by stepping the retarding potential on radial grid around the source, while the magnetic field was kept constant and adjusted to focus electrons of kinetic energy 2.2 keV. The measured and calculated resolution functions of the spectrometer with resolution of 17 eV (FWHM) are depicted in Fig (**16**). The detector, a proportional counter with a resistive anode wire, had a position resolution of 1–2 mm. Although the source was cooled down to -50°C and a large liquid nitrogen trap was installed within the spectrometer, detector background continuously grew, and typically after three weeks a decontamination of the setup from tritium was necessary. Test measurements with the 7 keV conversion electrons from $^{57}$Co decay did not reveal any deposition of a residual gas that would exceed 10 ng $cm^{-2}$ on a cooled source.

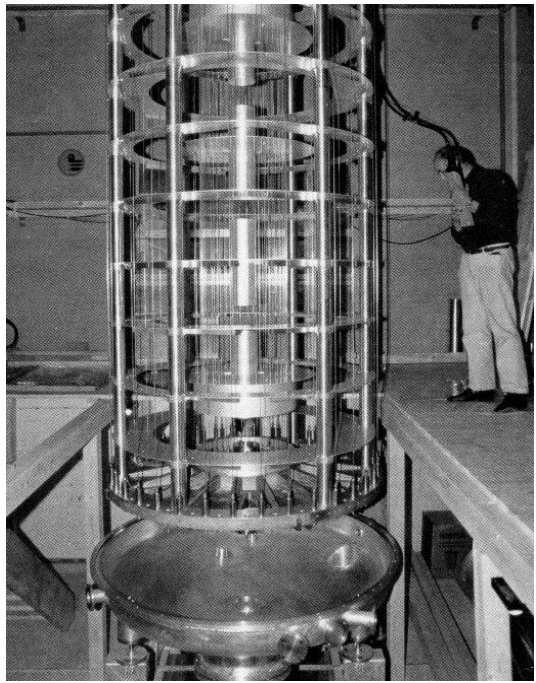

**Fig. (15)**. Toroidal magnetic spectrometer [111] of the Tretyakov type [109] installed at the Zurich University and used in several β-decay experiments. Distance between source and detector is 2.65 m. Photo reproduced from [124].

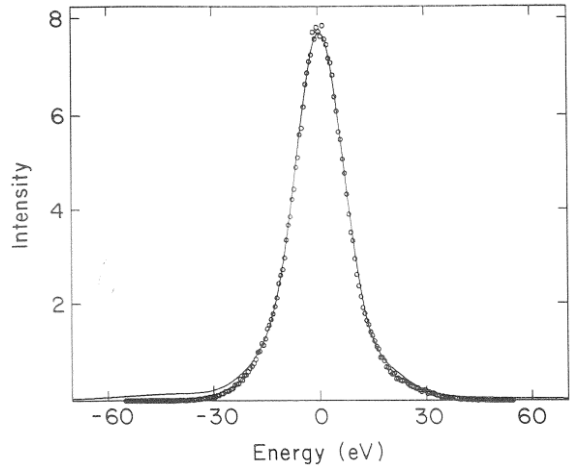

**Fig. (16).** Resolution function of the toroidal magnetic spectrometer at Zurich University [123]. The solid line with the resolution of 17.0 eV (FWHM) was derived from the measured $L_1$ conversion-electron line of the 30.898 keV transition in $^{195}$Au ($\Gamma$ = 6.25 eV). Points were obtained from a Monte Carlo simulation.

Stoeffl and Decman at the Lawrence Livermore National Laboratory also constructed a toroidal β-ray spectrometer with a gaseous tritium source [112]. The quality of the equipment, similar to that of Robertson *at al* [110], was demonstrated by conversion electron spectra emitted by multiply charged $^{83m}$Kr in gaseous state. However, their tritium β-spectrum showed an anomalous structure near the endpoint yielding an unphysical result $m_{\nu_e} = -130 \pm 20$ eV.

Two groups at the Institute of Nuclear Research at Troitsk near Moscow and the Physics Institute of the University in Mainz independently developed a new type of an integrating electron spectrometer well suited for measurements of tritium β-spectra, since it can reach an energy resolution of units of eV while keeping an input solid angle of tens percent of 4π [125], [126], [127]. The instrument combines a retarding electrostatic field with a guiding inhomogeneous magnetic field (the so called MAC-E Filter). Contrary to common differential spectrometers, the resolution function of this device has no high-energy tail, a great advantage in the search for a signature of massive neutrino in the uppermost part of β-spectrum. Weinheimer *et al.* [128] employed the Mainz MAC-E-Filter to measure β-spectrum of molecular $T_2$ frozen on an aluminum substrate cooled down to 2.8 K. The spectrometer was set to the full width (FW) resolution of 4 eV near the tritium endpoint. The fresh source of $10^8$ Bq activity contained about 40 monolayers of $T_2$ as determined by ellipsometry. Mass spectrometry revealed about 30 % of protium contamination, probably originating from previous storage of tritium in a stainless steel container. There was a substantial escape of tritium from the source with an effective half-life of one weak. Therefore the data were taken for about ten days per source. The electrons were detected with a segmented silicon detector with a resolution of 2.0 keV FWHM at 20 keV energy. Data were stored event-by-event, allowing the checking of the distribution of time differences between successive events. Observed sudden increases in the count rate, probably triggered by microsparks in the spectrometer, were rejected. The remaining data were in accord with the statistics. An observable β-spectrum already emerged clearly from the background at 20 eV below the endpoint but the deduced $m_{\nu_e}^2$ shifted to increasingly more negative values with widening fit interval. Analyses of the last 137 eV of the tritium β-spectrum yielded an upper limit of $m_{\nu_e} < 7.2$ eV at 95 % confidence level. The final results of the Mainz and Troitsk experiments are discussed in Sect. 3.4.

The experimental results disproving the ITEP claim [117] are summarized in Tab. (**2**). To facilitate the comparison we present the $m_{\nu_e}$ upper limits calculated according to the unified approach [130] with quantities constrained to be non-negative. Undoubtedly, the case of the 30 eV-neutrino considerably improved β-spectroscopic methods aiming at the neutrino mass determination. In the words of the Nobel prize-winner V. L. Ginsburg, "Sometimes science is better served by a wrong result if it is published than withheld" (quoted in [124]).

**Table 2. Final results of the tritium β-spectra measurements exploring the 17 < $m_\nu$ < 40 eV claim of ref. [117]**

| Experiment | Spectrometer | Source | $m^2_\nu$ (eV$^2$) | $m_\nu$ (eV) at 90% CL |
|---|---|---|---|---|
| Kawakami et al.[122] Tokyo, 1991 | Magnetic, π√2 | Solid; cadmium salt of tritiated $C_{20}H_{40}O_2$ | $-65 \pm 85_{stat} \pm 65_{syst}$ | < 11 |
| Robertson et al. [110] Los Alamos, 1991 | Magnetic, toroidal | Gaseous tritium molecules | $-147 \pm 68_{stat} \pm 41_{syst}$ | < 6 |
| Holzschuh et al. [123] Zurich, 1992 | Magnetic, toroidal | Solid; tritiated octadecyltrichlorosilan | $-24 \pm 48_{stat} \pm 61_{syst}$ | < 10 |
| Weinheimer et al.[128] Mainz, 1993 | Electrostatic retardation with magnetic collimation | Solid, frozen tritium Molecules | $-39 \pm 34_{stat} \pm 15_{syst}$ | < 6 |
| Sun Hancheng et al. [129] Beijing, 1993 | Magnetic, π√2 | Solid; tritiated $C_{14}H_{15}T_6O_2N_3$ | $-31 \pm 75_{stat} \pm 48_{syst}$ | < 11 |

### 3.4 Recent measurements yielding $m_{\nu_e}$ < 2 eV

The best model-independent upper limits of the effective neutrino mass $m_{\nu_e}$, as defined by Eq. (5), originate from two measurements of the tritium β-spectrum using electrostatic spectrometers of the MAC-E-Filter type. Kraus et al. [38] improved the Mainz setup in the following way in comparison with its previous stage [128]. The shock-condensed, amorphous $T_2$ films were cooled to 1.8 K, which substantially slowed down the surface migration, leading to small crystals of size exceeding the average thickness of the $T_2$ layer. Unaccounted energy loss in these crystals formed at temperatures of 3–4 K was the main reason for previous negative values of $m^2_{\nu_e}$[128]. The source area and thickness were increased to 2 cm$^2$ and about 140 monolayers (~50 nm). Highly oriented pyrolytic graphite (HOPG) with low electron backscattering and atomic flat surface over wide terraces served as a source backing. A LHe-cooled chicane with inner carbon coating adsorbed traces of $T_2$ that escaped from the source. In this way, the source-dependent component of the detector background was eliminated. Electropolishing the spectrometer tank and its electrodes not only reduced outgassing and field emission but also removed tritium contamination from previous runs. The spectrometer was backed up to 330–420°C and conditioned up to ± 30 kV to further improve conditions of the experiment. Using this improved setup, 11 measurement series were carried out during five years. The following are examples of the effects considered in the β-spectra analysis: a continuously growing coverage of the source by 0.3 monolayers of $H_2$ per day; a $T_2$ loss of 0.17 monolayers/day obviously caused by the recoiling daughter molecules that each sputtered a few neighboring molecules from the source, and a self-charging of the $T_2$ film that reached about 2.5 V at the outer surface. Final states of the daughter molecule calculated by Saenz et al. [131] for gaseous $T_2$ were slightly modified for a solid $T_2$ and complemented with a prompt excitation of neighbors next to a decaying $T_2$ molecule. This effect (amounting to a surprisingly large value of ~5 % of all decays) is due to local relaxation of the lattice following the sudden appearance of a (T$^3$He)$^+$ ion. The final result of the Mainz neutrino mass experiment is $m^2_{\nu_e}$ = –0.6 ± 2.2$_{stat}$ ± 2.1$_{syst}$ eV$^2$, yielding $m_{\nu_e}$ < 2.3 eV at 95 % CL [38].

Aseev et al. [39] re-analyzed all data acquired by the Troitsk experiment in which the MAC-E Filter with the resolution of 3.7 eV (FW) at 18.6 keV was utilized to analyze β-particles emitted from a windowless gaseous source of molecular tritium (WGTS), similar to that of [110]. Details of the apparatus are described in [132]. The authors succeeded in suppressing the partial pressure of tritium in the spectrometer (of which the decay would cause an undesirable background) to about 10$^{-18}$ mbar. Superconducting solenoids provided a maximum magnetic field of 5 T, while the minimum field in the spectrometer analyzing plane was 1 mT. Electrons were detected with a Si(Li) detector with sensitive area of 2.3 cm$^2$. Due to the point-by-point character of the measurement, the intensity of the WGTS was checked every 15 minutes at the monitor point of β-spectrum at 18.0 keV where the counting rate was high. The hydrogen-isotope concentration in the WGTS was measured every two hours but nevertheless the main systematic error arose from an uncertainty of the tritium column density in the WGTS. Only runs with reliable experimental conditions were included in the re-analysis of β-spectra acquired during the 1994-2004 period. The former anomaly – a step-like structure of 5–15 eV below the endpoint with roughly 6 months period and an amplitude of ~6·10$^{-11}$ of all β-decays – was no longer observed. The final result of the Troitsk neutrino mass experiment is $m_{\nu_e}^2 = -0.67 \pm 1.89_{stat} \pm 1.68_{syst}$ eV$^2$, i.e. $m_{\nu_e} < 2.05$ eV at 95 % CL [39].

From these two independent experiments it is possible to calculate their weighted average, $m_{\nu_e}^2 = -0.64 \pm 1.95$ eV$^2$, which according to [130] gives the current upper limit of $m_{\nu_e} < 1.6$ eV at 90 % CL.

**3.5 Experiments under construction**
**3.5.1 KATRIN**

The neutrino mass experiments at Mainz [38] and Troitsk [39] reached their sensitivity limits. The successor, the international collaboration KATRIN (the Karlsruhe tritium neutrino experiment) [40] was founded in 2001 with the aim of increasing the sensitivity to $m_{\nu_e}$ by one order of magnitude up to 0.2 eV. In order to reach this improvement (increase by two orders of magnitude with respect to the experimental observable $m_{\nu_e}^2$) the collaboration developed an extremely large MAC-E-Filter of 23 m length and 10 m diameter to analyze the endpoint region of the tritium β-spectrum. A WGTS [133] held at 30 K with a stability better than 30 mK [134] will provide 10$^{11}$ β-particles per second. The necessary amount of chemically and isotopically purified tritium will be supplied by a unique tritium laboratory which is a part of the Karlsruhe Institute of Technology. The tritium purity will be monitored by laser Raman spectroscopy [135]. A system of superconducting magnets will guide β-particles in a magnetic flux tube of 190 Tcm$^2$ into a pre-spectrometer (a smaller MAC-E-Filter of moderate resolution). This instrument [136] will cut off the prevailing part of β-spectrum, not bearing any information about the neutrino mass. The remaining uppermost part will be analyzed by the main spectrometer with a resolution of 0.93 eV (FW). A strong differential and cryogenic pumping will ensure that the tritium pressure inside the spectrometer will be 14 orders of magnitude smaller than that in the middle of the WGTS. A system of wire electrodes with a more negative potential than the main spectrometer vessel will not allow secondary electrons produced by cosmic muons to penetrate into the spectrometer volume and increase background. Electrons will be detected by an array of 148-pixel Si PIN detectors [137]. A high voltage in the region of 18 kV will be measured with the help of a precision voltage divider [138]. In addition, the stability of the high-voltage will be continuously checked by the monitor spectrometer (reconstructed Mainz MAC-E-Filter) using conversion electrons from a solid $^{83}$Rb/$^{83m}$Kr source with energy stability of 0.3 ppm/month [139]. The challenge of controlling the electrostatic potentials in this experiment has been discussed in ref. [140].

The possibility of increasing the statistical KATRIN sensitivity to $m_{\nu_e}$ by implementing the time-of-flight method has been explored in ref. [141]. A KATRIN chance to see an admixture of sterile neutrinos is discussed in Sect. 3.7.2. Further details about the KATRIN set-up can be found in the design report [142] and in a recent review [31]. An example of a simulated β-spectrum is shown in Fig. (**17**). After five years of data taking the KATRIN sensitivity to $m_{\nu_e}$ should reach 0.2 eV at 90 % CL if no mass signal is observed, while the $m_{\nu_e}$ mass of 0.35 eV should be proven with 5σ significance.

**3.5.2 MIBETA, MARE and HOLMES**

The problem of electron energy losses within a radioactive source and complications concerning final states after a β-decay could be eliminated, at least in principle, by means of low-temperature microcalorimeters. When an investigated radionuclide is fully contained in a microcalorimeter absorber all energy released in a β-decay (excluding that of the neutrino) is transferred into a heat impulse measurable with a sensitive thermometer. The previous investigations focused on $^{187}$Re → $^{187}$Os + e$^-$ + $\bar{\nu}_e$ decay with a lowest known endpoint energy of 2.44 keV. In the $m_{\nu_e}$–sensitive part of the β-spectrum there are 400 times more β-decays of $^{187}$Re than in a tritium spectrum but $^{187}$Re half-life of 4.3· 10$^{10}$ y eliminates any usual β-spectroscopic techniques. Microcalorimeters measure a complete β-spectrum and thus – contrary to magnetic and electrostatic instruments – they cannot select only a small subsection of β-spectrum sensitive to $m_\nu$.

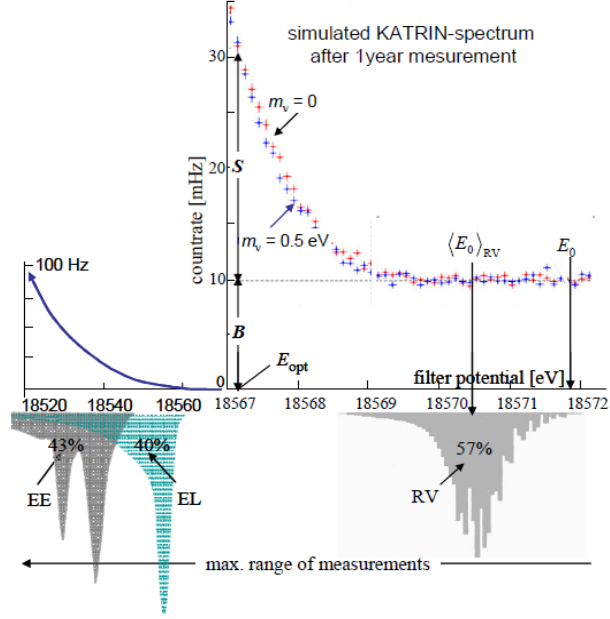

**Fig. (17).** Upper right: Simulated KATRIN data in the last 5 eV below the endpoint $E_0$ after 1 year of measurement for neutrino mass 0 eV (upper read crosses) and 0.5 eV (lower blue crosses). Upper left: Schematic increase of the count rate towards the lower energy of β-particles. Lower half: EE and RV: Electronic and rotational/vibrational excitation of the daughter molecule $(T^3He)^+$ occurring with 43 % and 57 % probability, resp.; EL: energy loss of 18.5 keV beta's within $T_2$-gas, affecting about 40 % of them. The figure is reproduced from ref. [140].

In order to avoid a pile-up effect[14] deforming the shape of the $^{187}$Re spectrum, its activity in a single microcalorimeter had to be limited to roughly 1 Bq.[15] So far the best results were obtained in the MIBETA (Milano β-ray) experiment [143] with eight monocrystals of $AgReO_4$ of total mass of 2.2 mg. The average resolution of individual microcalorimeters was 25 eV at 2.5 keV and the time resolution (for the impulse rise time) was about 500 μs. Every two hours of taking the β-spectrum were followed by 20 minutes of energy-calibration with soft X-ray lines. During 4000 hours $6.2 \cdot 10^6$ β-particles with energy above 700 eV were registered and their Kurie plot is shown in Fig. (**18**). The final result of the MIBETA experiment was $m_{\nu_e}^2 = -112 \pm 207_{stat} \pm 90_{syst}$ eV$^2$, yielding $m_{\nu_e} <$ 15 eV at the 90 % CL. In the energy interval between 470 eV and 1.3 keV, an oscillation modulation of the data due to the β-environmental fine structure effect in $AgReO_4$ was seen. This phenomenon, caused by a local molecular or crystal environment of the β-decaying nucleus, was first observed in metallic rhenium [144].

In order further to exploit possibilities of the calorimetric method, the MARE (Microcalorimeter arrays for a neutrino mass experiment) collaboration was founded in 2005 with the aim to achieve a neutrino mass sensitivity of 0.2 eV [145]. According to simulations, about $10^{14}$ β-decays of $^{187}$Re have to be detected. The final set-up was planned to consist of 50 000 microcalorimeters, each of them exhibiting 5 eV FWHM energy resolution with a 1 μs time resolution. Matrixes, of typically ~5000 microcalorimeters and of relatively small size, were intended to operate either in existing cryogenic apparatuses or those constructed among other uses for the MARE purpose. The major challenge remained coupling the metallic rhenium absorber to a low-temperature sensor [146].

The MARE collaboration therefore examines the electron capture (EC) decay $^{163}$Ho + $e^- \rightarrow$ $^{163}$Dy + $\nu_e$ with $T_{1/2}$ = 4570 y [147]. Due to a low-energy $Q_{EC}$ of about 2.5 keV, EC is allowed only for the M, N and O atomic shells. This approach to the determination of $m_{\nu_e}$ had already been suggested in 1982 [148] (see also up-date [149])[16]. Low-temperature microcalorimeters are well suited to detect all the energy released in the EC decay, except for the energy taken off by the neutrino. The endpoint region of the calorimetric spectrum is sensitive to the neutrino mass in a similar way to how it is in the β-decay. In addition, this spectrum consists of Lorentz peaks (LP)$_i$ with parameters relating to a particular atomic subshell $i$ taking part in the EC decay. More explicitly,

$$\frac{dN}{dE_c} \sim (Q_{EC} - E_c)\,[(Q_{EC} - E_c)^2 - m_\nu^2]^{1/2} \times \sum_i (LP)_i, \qquad (14)$$

where $Q_{EC}$ is the mass-difference of the mother and daughter atoms and $E_c$ is the calorimetric energy [148].

---

[14] Two or more events that occur in the same detector within a time interval shorter than the signal rise-time are recorded as one false event contributing to an unresolved pile-up background spectrum.
[15] This is the activity of about 1 mg of rhenium with a 63% natural isotopic abundance of $^{187}$Re.

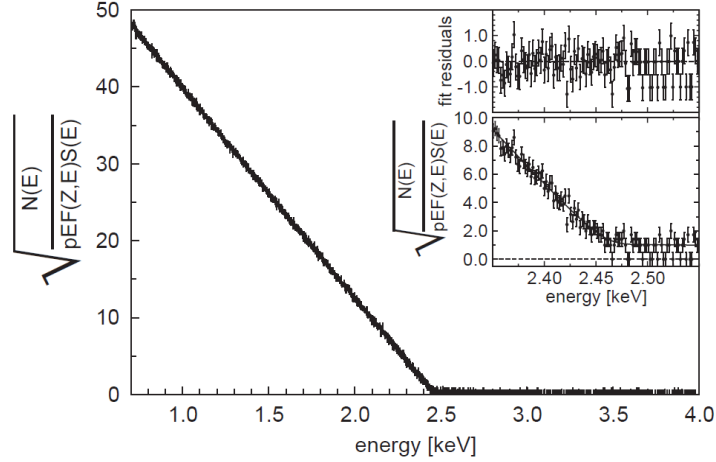

**Fig. (18).** Kurie plot of the $^{187}$Re ($5/2^+$) → $^{187}$Os ($1/2^-$) β-decay measured with an array of low-temperature microcalorimeters [143]. The least-squares fit of the theoretical spectrum for the first forbidden unique transition yielded a value for $\chi^2$ per degree of freedom of 0.905. The insert shows the fit of the uppermost part of the β-spectrum and the residuals of the fit in units of a standard deviation.

The status of the MARE experiment with $^{187}$Re and $^{163}$Ho isotopes was reported at the TAUP 2013 conference [147]. The first array of MARE-1, containing 31 thermistors equipped with AgReO$_4$, was assembled. At a working temperature of 85 mK, the 1.5 keV calibrating X-rays were recorded with an energy resolution of 28 eV and an impulse rise time of about 1 ms. The R&D in this field was recently supported by a new project HOLMES [152] that plans to deploy a 1000-channel multiplexed array of temperature edge sensors coupled to Au absorbers. Each of the absorbers will be implanted with $6.5 \cdot 10^{13}$ holmium nuclei yielding 300 decays of $^{163}$Ho per second. The detectors should exhibit energy resolution of ~1 eV and resolution time of ~1 μs. The plan is to collect $3 \cdot 10^{13}$ decay events during 3 years of measuring time. The aim of the HOLMES experiment is to reach a $m_\nu$ sensitivity of 0.4 eV and explore a possibility to enhance it to 0.1 eV.

The statistical sensitivity of $^{163}$Ho electron-capture neutrino mass experiments has been explored in ref. [151] by means of Monte Carlo simulations, see Fig. (**19**). Robertson [153] pointed out that the theoretical spectrum is not yet sufficiently understood to allow an eV-scale determination of the neutrino mass from a measured $^{163}$Ho spectrum. The reason is a complicated structure of energetically allowed final states of the dysprosium atom after the $^{163}_{67}Ho \rightarrow {}^{163}_{66}Dy$ decay. Very recently Faessler performed detailed calculations of the process using the relativistic Dirac-Hartree-Fock electron wave functions and concluded that the two-hole [154] and three-hole [155] excitations do not complicate the determination of the neutrino mass compared to the situation with one-hole states only.

---

[16] The best upper limit for the neutrino (not antineutrino) rest mass obtained using this method is $m_\nu <$ 225 eV at 95 % CL [150]. It was derived from the inner bremsstrahlung spectrum emitted by the 4p and 5p atomic electrons during the $^{163}$Ho decay. The measurements were carried out with a 30 mm$^2$ Si(Li) detector having resolution $\Delta E_{FWHM}$ = 137 eV at 2 keV and luminosity of 4 % of 4π. The source contained $3.3 \cdot 10^{17}$ atoms of $^{163}$Ho providing an activity of 1.6 MBq.

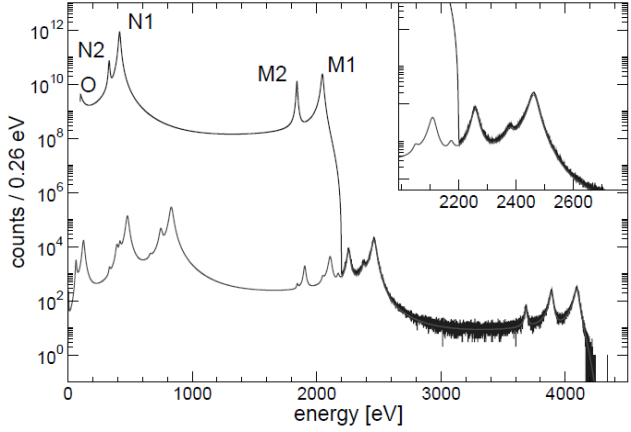

**Fig. (19).** A Monte Carlo simulation of the calorimetric spectrum of the $^{163}$Ho electron capture decay [151]. Calculations were based on the following assumptions: a decay energy $Q_{EC}$ = 2200 eV, a neutrino mass $m_\nu$ = 0, a total number of decay events $N_{ev}$ = $10^{14}$, a microcalorimeter energy resolution $\Delta E_{FWHM}$ = 2 eV and a pile-up fraction $f_{pp}$ = $10^{-6}$. The bottom curve is a fit of the pile-up spectrum. The insert shows the uppermost part of the spectrum depending on the neutrino mass in a similar way as in the case of the β-ray spectra. Note the complexity of the pile-up spectrum at the end of the $^{163}$Ho decay spectrum.

**3.6 New approaches**
**3.6.1 ECHo**

The ECHo collaboration (the electron-capture $^{163}$Ho experiment) [156] intends to investigate the electron neutrino mass by low temperature metallic magnetic calorimeters (MMC)[17] that involve gold absorbers with implanted $^{163}$Ho ions [157]. A first prototype chip consisted of four pixels each having two gold layers of dimensions (190 ×190 × 5) μm$^3$. $^{163}$Ho ions were implanted on an area (160×160) μm$^2$ of the first layer while a second gold layer was put on the top of the first one [156]. A spectrum of $^{163}$Ho corresponding to an energy resolution $\Delta E_{FWHM}$ ~12 eV and a rise-time of about 100ns is shown in Fig. (20).

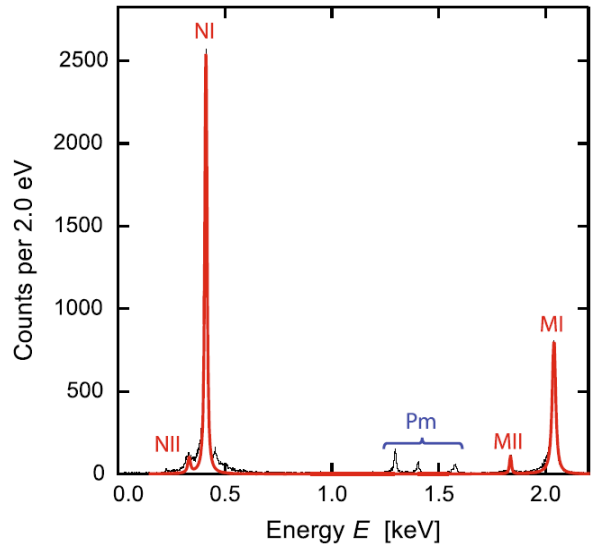

**Fig. (20).** A spectrum of the $^{163}$Ho electron-capture decay measured by a low temperature calorimeter [156]. The fitted lines are shown in red. Lines labeled Pm correspond to the EC decay of a small amount of $^{144}$Pm admixed in early stage of the $^{163}$Ho measurement.

---

[17] The temperature sensor of the MMC is a paramagnetic alloy (e.g. a dilute alloy of erbium in gold) placed in a small magnetic field. The temperature rise, caused by absorption of low energy photons and electrons after the EC decay, leads to a change of magnetization of the sensor registered as a change of magnetic flux by a sensitive SQUID magnetometer.

A first 64-pixel chip with an integrated micro-wave SQUID multiplexer was also developed. The ECHo collaboration hopes to reach an energy resolution below 5 eV for multiplexed MMC detectors with implanted $^{163}$Ho. A great effort is devoted to production of several MBq of chemically and isotopically pure $^{163}$Ho. In order to reach sub-eV sensitivity on $m_{\nu_e}$ a total statistics of at least $10^{14}$ events in the full calorimetric spectrum of $^{164}$Ho EC spectrum will be necessary. The ECHo collaboration intends to determine the $^{163}$Ho decay energy $Q_{EC}$ with a precision of 1 eV or better using a novel Penning-trap mass spectrometer PENTATRAP [158]. It needs to be verified that a de-excitation of the daughter $^{163}$Dy is not accompanied by unforeseen metastable states that have a half-life longer than the ~0.1 μs rise-time of the present MMC. Recently, M. W. Rabin (quoted in [149]) warned that $^{163}$Ho atoms in a low-temperature microcalorimeter should be all bound in one type of chemical neighborhood otherwise the calorimeter would act as a sum of detectors with various values of $Q_{EC}$.

The efforts to improve calorimetrical measurements of the $^{163}$Ho EC spectra, including production and purification of the $^{163}$Ho source, were recently reviewed in ref. [159].

### 3.6.2 Project 8

Monreal and Formaggio [160] suggested a new technique for measuring the neutrino mass by precision spectroscopy of coherent cyclotron radiation emitted by tritium decay electrons in a magnetic field.
The angular frequency $\omega = 2\pi f$ of this radiation depends on the kinetic energy $E$ of the electron but not on the angle between the electron velocity and direction of the magnetic field vector $B$ hence

$$\omega = \frac{eB}{m_e\left(1+\frac{E}{m_e c^2}\right)} \quad (15)$$

where $e$ and $m_e$ are the electron charge and rest mass, respectively. This method differs principally from the previous ones based either on an analysis of electron movement in magnetic and/or electric fields or on calorimetric measurement of energy released by electrons and photons in a β-decay. The proposed method takes advantage of an extraordinary precision of frequency measurements. Contrary to taking the energy spectra point-by-point as usual, but similarly to calorimetric measurements, all frequencies are recorded simultaneously. Therefore, precautions have to be taken against the pile-up effect that would otherwise spoil the measured β-ray spectra. The systematic errors occurring in the frequency measurements differ substantially from those in previous techniques, which is an advantage in the effort to achieve a reliable neutrino mass determination.

The authors [160] considered a reference design where an electron with energy $E$ = 18.575 keV emitted from the endpoint region of the tritium β-spectrum spirals in a magnetic field of $B$ = 1 T, with a radius of cyclotron orbit smaller than 0.5 mm. According to Eq. (15) the electron emits cyclotron radiation of frequency $f$ = 27.01090483 GHz (wave-length of about 1.1 cm), which is well within the range of commercially available radio-frequency antennas and detectors. The total power of the cyclotron radiation depends on $B^2$, $E$, and angle $\theta$ of the electron direction relative to the field vector. For the considered case, the irradiated power reaches 1.2 fW which should be sufficiently high to enable single-electron detection. In order to achieve resolution of $\Delta E_{instr}$ = 1 eV, a relative precision of $\Delta f/f = 2 \cdot 10^{-6}$ is necessary and the frequency signal should be monitored for at least $t_{min}$ = 38 μs. The energy radiated by an electron during this time will not exceed 0.35 eV. In order not to disturb the electron motion by a $T_2$ - $e^-$ inelastic scattering the density of molecular tritium should not exceed $1.4 \cdot 10^{11}$ cm$^{-3}$. In the actual experiment the narrowband signal of frequency $f$, as recorded by a stationary antenna, will be more complicated, mainly due to a Doppler shift.

The Project 8 collaboration [161] intends to develop a source of atomic tritium in a magnetic configuration that would trap both spin-polarized atoms and β-particles. This would avoid a rotational-vibrational broadening of 0.36 eV that accompanies the β-decay of molecular tritium. The source temperature, now limited to 30 K in $T_2$, could be lowered to 1 K thus reducing the Doppler-shift broadening. Preliminary results of the potential sensitivity of this novel method are exhibited in Fig. (**21**). While attractive, they should be regarded as the best estimate that could be reached with this approach [161]. The final goal of Project 8 is to reach a sensitivity to the effective neutrino mass $m_{\nu_e}$ (Eq. 5) as low as 0.05 eV. If this experiment were to reliably prove that $m_{\nu_e}$ is less than this limit, it would imply that the hierarchy of neutrino mass states is normal one, i.e. $m_1 < m_2 < m_3$.

Using conversion electrons emitted by $^{83m}$Kr in a gaseous state, the Project 8 collaboration recently proved that cyclotron radiation can detect and measure the energy of a single electron [162]. It is interesting to see in Fig. 2 of ref. [162] repeated cycles of a continuous decrease of the electron kinetic energy by emission of cyclotron radiation followed by a sudden jump caused by a collision with residual gas.

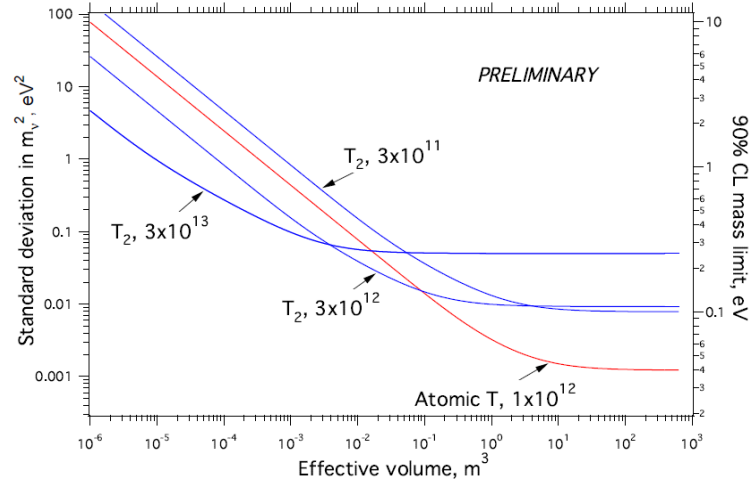

**Fig. (21).** The neutrino mass sensitivity expected in the Project 8 measurement using cyclotron radiation of single electrons emitted in tritium β-decay [161]. The dependence on the effective volume of gaseous sources of molecular or atomic tritium is shown for indicated number densities per cm$^3$. A source temperature will be 30 K for molecular T$_2$ and 1 K for atomic tritium. Further assumptions were a magnetic field uniformity of 0.1 ppm rms, a background of 10$^{-6}$ s$^{-1}$ eV$^{-1}$ and a live time of 3·10$^7$ s. The $m_{\nu_e}$ sensitivity for T$_2$ is limited to about 100 meV because of the width of the rotational-vibrational states of the ($^3$HeT)$^+$ daughter ion and a minimum operating temperature near 30 K.

**3.6.3 Further proposals**

In this section we mention a few additional approaches for measurements of the neutrino mass. However, the feasibility of these experiments still awaits experimental verification.

*The bound-state β-decay of tritium*

Cohen *et al.* [163] considered a possibility of determining the neutrino mass by examining the bound-state β-decay of tritium. In this two-body process, the electron is born into an unoccupied bound state of the daughter atom (preferably in the K-shell) and the available decay energy is divided between a monoenergetic neutrino and the recoiled daughter nucleus[18]. According to calculations [165] the branching ratio of bound-state to continuum β-decays of atomic tritium equals 0.69 %. About 80 % of the bound-state decays goes into electron creation in the 1s ground state of the $_2^3He$ neutral atom. The proposed method requires precision measurement of the change $\Delta V_R$ of the velocity of recoiling atoms due to the finite neutrino mass since [163]

$$\frac{\Delta V_R}{V_R} = \frac{1}{2}\left(\frac{m_{\nu_e}c^2}{Q_\beta}\right)^2 \qquad (16)$$

For the 0.2 eV neutrino mass sensitivity expected in the tritium β-decay experiment KATRIN [40] $\Delta V_R/V_R$ amounts to 5.6·10$^{-11}$. This could probably be measured using resonant excitation by a laser beam with its frequency tunable around an atomic spectral line of $^3$He. The source of atomic tritium should be kept at very low temperature to suppress the thermal motion of the atoms that smears out the velocity distribution. Background caused by $^3$He$^+$ ions produced in a dominant three body β-decay could be reduced by removing the ions using an electric field.

---

[18] The bound-state β-decay may become important for completely ionized heavy atoms at high temperatures in stars. It was observed for the first time by storing bare $_{66}^{163}Dy^{66+}$ ions in a heavy-ion storage ring [164].

*Radioactive ions in a storage ring*

A possibility of measuring the neutrino mass with radioactive ions in a storage ring was explored by Lindroos *et al.* [166]. This novel kinematic (and thus model-independent) approach requires a tuning of an ion beam that causes most of β-particles to move forward with respect to the beam while those from the endpoint region of the β-spectrum move backwards. If the ratio of backward to forward moving electrons could be determined with sufficient precision the proposed method would yield information on $m_{\nu_e}$. The idea is to store a beam of β-radioactive ions with a very low boost factor (e.g. $\gamma = 1.037$ for $^3$H ions) in an evacuated chamber with a weak magnetic field parallel to the beam line. A detector on the forward wall of the chamber will monitor the huge number of forward moving β-decays while another detector on the backward wall will count a small number of electrons spiraling backwards in the laboratory frame. In this way β-particles from the neutrino mass-sensitive endpoint region can be separated from the prevailing β-spectrum part. In order to reach or exceed the 0.2 eV sensitivity for $m_{\nu_e}$ a β-emitter with $Q = M - M' - m_e$ of a few keV (here $M$ and $M'$ are the masses of the mother and daughter ions) should be identified[19] and the number of detected β-particles should be at least $10^{18}$. The experimental requirements are illustrated in Fig. (22). Major challenges of the proposed experiment are the separation of the forward and backward moving electrons with a precision of order $10^{-16}$ while keeping the ion momentum spread $\delta p_I/p_I$ in the range $10^{-4} - 10^{-5}$, and avoiding any scattering of electrons or beam ions that would increase background in the backward detector.

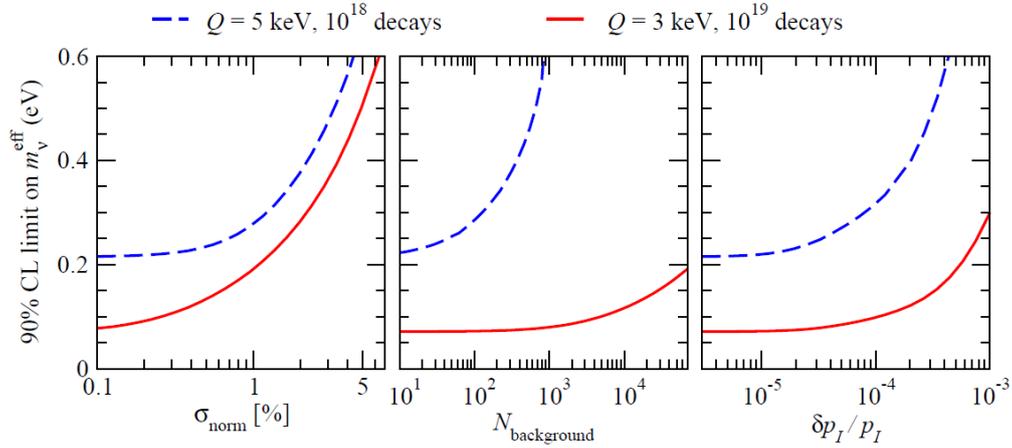

**Fig. (22).** Sensitivity to the neutrino mass at the 90 % CL predicted for the measurement of radioactive ions in a storage ring [166]. ($m_\nu^{eff}$ in the figure is identical with $m_{\nu_e}$ defined in Eq. (5) of this work.) Two β-emitters with energies $Q = 5$ and 3 keV, and total amount of $10^{18}$ and $10^{19}$ recorded decay events, respectively, are considered. Three effects influencing $m_{\nu_e}$ were examined: a normalization uncertainty $\sigma_{norm}$ on the beam flux (left panel), background $N_{background}$ in the backward detector (middle panel) and a momentum spread $\delta p_I/p_I$ of the initial ion (right panel).

*β-transitions to an excited state of a daughter nucleus*

As can be seen in Fig. (5), a finite value of the neutrino mass appreciably influences the β-spectrum only in its uppermost narrow region comparable with $m_{\nu_e}$. There might be β-transitions to an excited state of the daughter nucleus exhibiting extremely low decay energy $Q_\beta$ and thus a high sensitivity to $m_{\nu_e}$. Kopp and Merle [167] discussed several candidate isotopes undergoing $\beta^\pm$, bound state β, or electron-capture decay. The authors also showed that partial ionization of the parent atom could help tune $Q_\beta$ values to $\ll 1$ keV since every spectator atomic electron contributes to $Q_\beta$ with its energy gain or loss caused by the change of the nuclear charge during the β-decay. An example of a $^{194}_{77}Ir \rightarrow ^{194}_{78}Pt$ $\beta^-$ decay with $Q_\beta = 2246.9 \pm 1.6$ keV and $T_{1/2} = 19$ h is shown in Fig. (23).

---

[19]Unfortunately, $^{187}$Re with $Q = 2.4$ keV is not suitable for measurements of this type due to its half-life of $4.0 \cdot 10^{10}$ y.

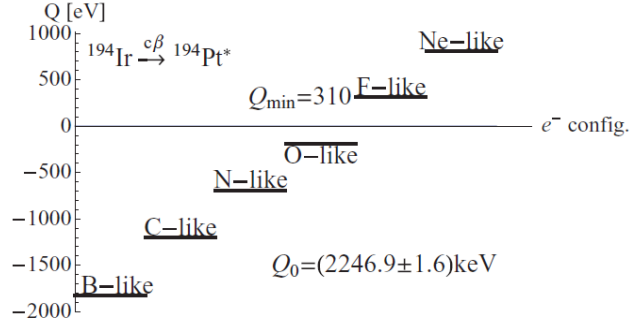

**Fig. (23).** A change of the $^{194}_{77}Ir \rightarrow ^{194}_{78}Pt$ decay energy by ionization of the parent iridium atom [167]. The figure demonstrates the magnitude of changes but, due to a large uncertainty in the decay energy, it is only illustrative.

However, the number of stored parent atoms needed for achieving the expected $m_{\nu_e}$ sensitivity of the KATRIN experiment is many orders of magnitude beyond the capabilities of present ion traps and storage rings. In order to identify suitable candidates for measurements of ultralow energy β-transitions to excited daughter states more precise data on $Q_\beta$ and excitation spectra of daughter nuclei are needed.

*β-decay of ultracold atomic tritium*

Jerkins *et al.* proposed a determination of the electron antineutrino mass $m_{\nu_e}$ from a complete kinematic reconstruction of the tritium β-decay [168]. The authors expect that determination of kinetic energy and three components of the momentum of both β-particle and recoiled $^3$He$^+$ ion in each of the recorded β-decay events, carried out with sufficient precision, will lead to a higher sensitivity for $m_{\nu_e}$ than measurement of the shape of the β-spectrum alone. The principle of their experimental setup is shown in Fig. (**24**). An assembly of $10^{13}$ tritium atoms magnetically trapped in several spheres of 100 μm in diameter and cooled to a temperature below 1 μK should be observed by three detectors: (i) a microchannel plate (MCP) of the size 15 × 15 cm$^2$ with a spatial resolution of 2 μm and a timing resolution of 20 ps, (ii) two optical lattices of the size of 10 × 10 × 1 cm$^3$ each, containing rubidium atoms highly excited to the 53s Rydberg states, (iii) a hemispherical electrostatic analyzer exhibiting a resolution of 5 meV for β-particles pre-retarded by 18.1 kV. Full kinematic information about the $^3$He$^+$ ion shall be derived from *x* and *y* coordinates of the hit position and time-of-flight of the ion. The innovative concept of an optical lattice filled with rubidium atoms should allow measurement of two components of the β-particle momentum almost without altering its kinetic energy.

Numerical simulations [168] indicate that the proposed method could reach an $m_{\nu_e}$ sensitivity of 0.2 eV, the same as the one intended for the KATRIN experiment [40]. However, Otten argued in his comment [169] on the Jerkins proposal [168] that a factor of $10^6$ is lacking in necessary counting statistics. Jerkins *et al.* opposed [170] that their extraordinary high sensitivity is a consequence of suitable correlation between the shape of the β-spectrum and the shape of the neutrino mass-squared peak. Otten also demonstrated that spurious electromagnetic fields, decreasing the $m_{\nu_e}$ sensitivity, need to be controlled at almost an inconceivable level; the same holds for checking the stability and spatial distribution of the retarding potential [169]. It is known that hemispherical electrostatic analyzers with narrow input and output slits are capable of a high energy resolution[20] but their rather low luminosity was not considered in the Jerkins proposal, as was also reminded in [169]. A verification of this new concept for the $m_{\nu_e}$ determination will have to wait for a development of sources of ultracold atomic tritium[21] as well as for Rydberg track chambers.

---

[20] For example, the electron energy analyzer for photoelectron spectroscopy measured the Xe5p$_{3/2}$ spectrum in the gas phase with the resolution better than 2.7 meV at pass energy of 2 eV [171].
[21] Even temperatures much lower than 1 μK would be desirable to decrease the dominating spread in the values of $m_{\nu_e}^2$, see Fig. (**3**) in [168].

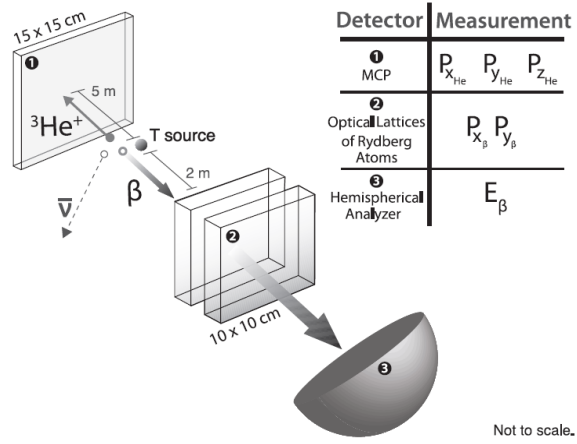

**Fig. (24).** Diagram illustrating the mechanism of the β-decay experiment proposed for the kinematic reconstruction of the neutrino mass from coincident measurement of β-particles and $^3He^+$ ions emitted by ultracold atomic tritium [168].

*Macro-coherently amplified radiative emission of a neutrino pair*

For completeness, we describe one more example that is not directly connected with β-ray spectroscopy. Fukumi *et al.* [172], Dinh *et al.* [173] and Tashiro *et al.* [174] explored theoretically a new type of atomic or molecular spectroscopy that could yield – at least in principle – the $m_i$ values of individual neutrino mass states and possibly determine other yet unknown neutrino properties. The basic idea is to prepare a collection of excited atoms or molecules in a metastable state |e> and then induce their cooperative and coherent deexcitation to a ground state |g>. The authors plan to measure spectra of low energy photons γ in the extremely rare process |e> → |g> + γ + $v_i v_j$ where $v_i$ and $v_j$ (with i, j = 1, 2, 3) denote the neutrino mass states. The process, called radiative emission of a neutrino pair (RENP) is a combination of second-order quantum electrodynamics (proceeding via virtual intermediate atomic or molecular states) and weak interaction. The released energy, about 8 eV in a Xe atom [172], about 2 eV in a Yb atom [173] and about 0.8 eV in an $I_2$ molecule [174], is much closer to the expected neutrino masses than in a nuclear β-decay. If the number density of coherently deexciting atoms or molecules will approach the Avogadro number per $cm^3$, the yet unobserved RENP rate may become measurable. Then in the continuous photon spectrum six threshold locations should appear at

$$\hbar\omega_{ij} = \frac{\epsilon_{eg}}{2} - \frac{(m_i c^2 + m_j c^2)^2}{2\epsilon_{eg}}, \qquad (17)$$

as shown in Figs. (**25**) and (**26**). $\epsilon_{eg}$ is the energy difference between the initial and final state. As a first step, Fukumi *et al* [172] plan to prove the macro-coherence in the twin process |e> → |g> + γ + γ. The RENP free of quantum electromagnetic backgrounds was treated theoretically in ref. [175]. Calculations [176] proved that a possible existence of one additional mass state of the light sterile neutrino (discussed in Sect. 3.7) would not decrease the sensitivity of the RENP method to active neutrinos.

### 3.7 Sterile neutrinos

Almost all experimental results in neutrino physics are consistent with the concept of three active neutrinos coupled to weakly interacting Z and $W^\pm$ bosons. Former claims for an admixture of a 17 keV neutrino have been rejected, see Sect. 3.7.1. Nevertheless, several short-baseline neutrino oscillation experiments and some cosmological observations indicate (with about 2–3σ significance) a possible existence of one or even two additional light sterile neutrinos. These so far hypothetical neutrinos would not have any ordinary weak interactions except those induced by mixing. Those alternatives are often referred to as the 3+1 and 3+2 scenarios. Experimental and theoretical aspects of light sterile neutrinos are treated in detail in ref. [10], [79], and the current experimental status and proposed searches of sterile neutrinos are listed in refs. [13], [177], [178].

In order to illustrate the present situation we quote a few recent examples: Archidiacono *et al.* [179] analyzed all available cosmological data assuming the 3+1 scenario and derived $m_{sterile} \approx m_4 = 0.44^{+0.33}_{-0.32}$ eV, while the analysis of short-baseline oscillation experiments yielded $m_4 = 1.27^{+0.15}_{-0.30}$ eV (on 2σ level) [180]. On the other hand, the Planck collaboration that measures the temperature anisotropy of the CMB [66], analyzed

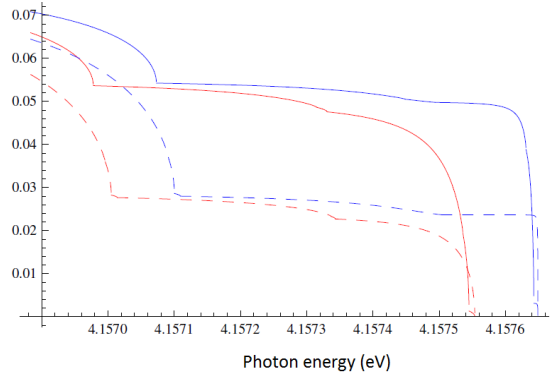

**Fig. (25).** Relative photon intensity in the threshold regions of the xenon RENP spectrum calculated for Dirac neutrinos assuming two different values of the smallest neutrino mass, 2 and 20 meV [172]. The solid lines correspond to the normal hierarchy ($m_1 < m_2 < m_3$) and the dashed curves to the inverted hierarchy ($m_3 < m_1 \approx m_2$) of the neutrino mass states.

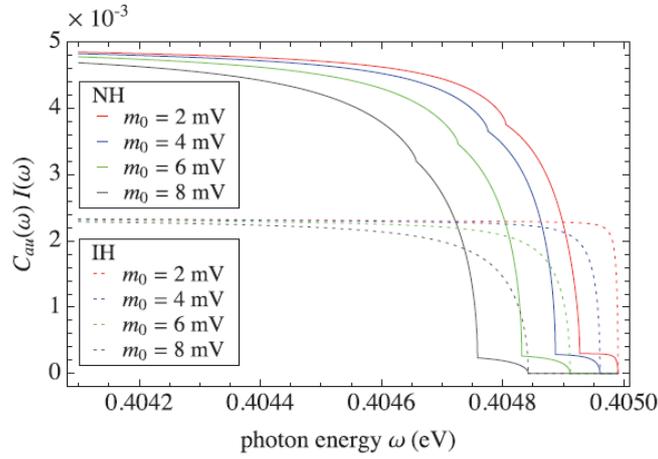

**Fig. (26).** Endpoint region of the photon RENP spectrum of the iodine molecule calculated for Majorana neutrinos with indicated values of the smallest neutrino mass $m_0$ [174]. Solid and dashed curves correspond to the normal (NH) and inverted (IH) hierarchy of neutrino mass states. The CP-conserving Majorana with both phases equal to zero was assumed

their data in combination with other cosmological surveys and found no clear indications for sterile neutrinos and derived $\sum m_i < 0.23$ eV [181]. An *et al.* [182] searched for a light sterile neutrino using data of the Daya Bay experiment on oscillation of reactor neutrinos. The measured data are in accord with the concept of three active neutrinos and lead to upper limits for $|U_{14}|^2$ in the region of $10^{-3}$ eV$^2 \leq |\Delta m^2_{41}| \leq 0.1$ eV$^2$. Abe *et al.* analyzed the Super-Kamiokande atmospheric neutrino data in the 3+1 framework assuming $|U_{e4}|^2 = 0$. No evidence of sterile neutrinos was observed. The following limits, $|U_{\mu 4}|^2 < 0.041$ and $|U_{\tau 4}|^2 < 0.18$ (90 % CL), were found for $\Delta m^2 > 0.1$ eV$^2$ [183]. Boyarsky *et al.* examined hypothetical keV sterile neutrinos as a possible component of the dark matter. He derived from astroparticle observations that an admixture of sterile neutrinos to the active ones should be less than $10^{-7}$ [184]. Leistedt *et al.* found that the concept of sterile neutrinos did not improve the concordance among various cosmological observations [185]. In the following two sections, we summarize former β-spectroscopic searches for sterile neutrinos and outline their perspectives.

### 3.7.1 Previous experiments

Already in 1980 Shrock examined a possibility for searching for additional neutrino mass states $m_i$ (i > 3) in β-ray spectra also [6]. He emphasized that an admixture of each of such states would produce a discontinuity in the β-spectrum slope similar to those shown in Fig. (**7**). From the position and amplitude of the $i$th kink the values of $m_i$ and $|U_{ei}|^2$ entering Eq. (5), could be determined. In principle, the signature of a specific mass state $m_i$ should be observable in the spectrum of every β-emitter with the endpoint energy $E_0$ at kinetic energy equal to $E_0 - m_i$. Shrock analyzed the experimental Kurie plots of a large number of β-decays. Taking into account that Kurie plots often deviate from linearity due to energy losses of low energy β-particles, he found a possible admixture $|U_{ei}|^2 \leq 0.1$ for $0.1$ keV $\leq m_i \leq 3$ MeV.

Schreckenbach et al. [186] searched for an admixture of sterile neutrinos in the β-decay of $^{64}$Cu ($T_{1/2} = 12.7$ h) that proceeds via β$^+$ and β$^-$ branches with endpoint energies of $E_{0\beta^+} = 653$ keV and $E_{0\beta^-} = 579$ keV, respectively. The $^{64}$Cu activity of $3.3 \cdot 10^{11}$ Bq was produced by (n, γ) reaction in a thermal neutron flux of $3 \cdot 10^{14}$ cm$^{-2}$ s$^{-1}$. A target of an area of (5 ×10) cm$^2$ and surface density of 500 μg cm$^{-2}$ was prepared by vacuum evaporation of 98% -enriched $^{63}$Cu onto an Al backing. The β$^+$ and β$^-$ spectra in the energy range 150-680 keV and 75-600 keV respectively were measured with a double-focusing magnetic spectrometer [187] equipped with a 32-wire proportional counter. The overall resolution of the set-up, including effects of the target thickness, was determined to be 1 keV at 300 keV using internal conversion lines in the reaction $^{63}$Cu (n, e$^-$). Analysis of the β$^+$ and β$^-$ spectra taken from the same source with the same instrument increased reliability of the analysis since the signature of a sterile neutrino with mass $m_4$ should appear in β$^+$ spectrum as well as β$^-$ spectrum at energies $E_{0\beta^+} - m_4$ and $E_{0\beta^-} - m_4$, respectively. An upper limit for the admixture of a neutrino state with a mass between 110 and 450 keV was determined to be 0.8% at the 90% CL.

Van Elmbt et al. [188] investigated the β-ray spectrum of $^{20}$F with $Q_\beta = 7.0$ MeV and $T_{1/2} = 11$ s. The $^{20}$F activity of 1.6 MBq was produced by the $^{19}$F (d, p) $^{20}$F reaction and measurements were done with an iron-free magnetic spectrometer set for a resolution of 3 keV at 4 MeV with a transmission of 0.1% of 4π. The electron energy loss within the LiF target of 280 μg cm$^{-2}$ surface density was 0.7 keV at 4 MeV. Reanalysis of the spectra revealed no admixture of sterile neutrinos, with an upper limit ranging from 0.59 to 0.18% for 400 keV $\leq m_4 \leq 2.8$ MeV [189].

Simpson continued in 1985 his investigation of the β-spectrum of tritium implanted into a Si(Li) x-ray detector [114] (see Sect. 3.2) and observed a distortion in the part of the spectrum below 1.5 keV that was consistent with an emission of a neutrino of mass about 17.1 keV and mixing probability of 3% [190]. Some of the subsequent measurements with semiconductor spectrometers, examining β-spectra of $^3$H, $^{14}$C, $^{35}$S and $^{63}$Ni, were consistent with about 1% admixture of the 17 keV neutrino, see e.g. Fig. (**27**). These claims stimulated several theorists to take such a mass state into account, see e.g. paper by Glashow [191].

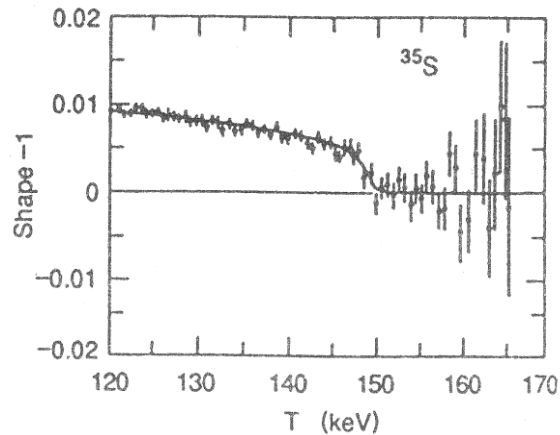

**Fig. (27).** The shape factor of the β-ray spectrum of $^{35}$S measured with a cooled Si(Li) detector [192]. In the energy region of 17 keV below the endpoint of $Q_\beta = 167$ keV, the observed spectrum deviates by 8 standard deviations from the expected single component β-spectrum. Solid line includes the 17 keV neutrino mass component with an admixture of $0.0084 \pm 0.0006_{stat} \pm 0.0005_{syst}$.

However, measurements with magnetic spectrometers were not consistent with the claim. In particular, an upper limit of the admixture of the 17 keV neutrino in the $^{63}$Ni decay was put as low as 0.073% at 95% CL [193]. The hypothesis of the 1% admixture was rejected by 22 σ, see Fig. (**28**). An upper limit for an admixture of neutrino with mass between 10.5 and 25 keV was found to be $|U_{e4}|^2 < 0.15\%$ at the 95% CL. An example of their β-spectrum of $^{63}$Ni is depicted in Fig. (**28**).

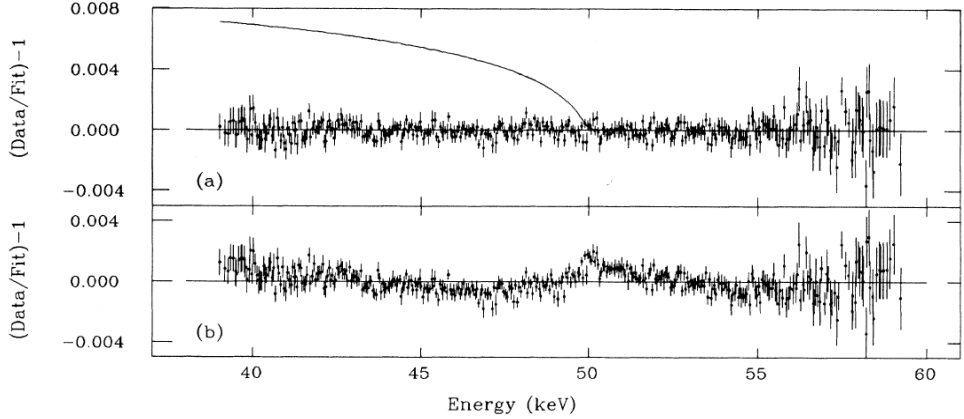

**Fig. (28).** The β-spectrum of $^{63}$Ni measured using a magnetic spectrometer compared with the best fit for $|U_{e4}|^2$ free (a) and $|U_{e4}|^2 = 1\%$ (b). The smooth curve in (a) illustrates the size of 1% effect of the 17 keV neutrino. The figure was reproduced from ref. [193].

The newer measurements with semiconductor spectrometers also did not support the 17 keV neutrino claim. For example, Mortara et al. [194] examined a β-spectrum of $^{35}$S ($Q_\beta$ = 167.2 keV, $T_{1/2}$ = 87.5 d) with an apparatus consisting of a superconducting solenoid that collected β-particles from the source to a high-resolution Si(Li) spectrometer. This arrangement exhibited an input solid angle of 50% of 4π and thus weak sources (typically less than 4 kBq) of better spectroscopic quality could be employed. The mixing probability of the 17 keV neutrino was determined to be –0.0004 ± 0.0008$_{stat}$ ± 0.0008$_{syst}$, in disagreement with several previous experiments using solid-state detectors. The authors proved sufficient sensitivity of their apparatus by introducing a known amount of $^{14}$C ($Q_\beta$ = 156.5 keV, $T_{1/2}$ = 5730 y) into their $^{35}$S source. Fitting the β-spectrum of the composite ($^{35}$S + $^{14}$C) source with a pure $^{35}$S spectrum shape led to the reduced $\chi^2$ value of 3.5 while fitting with a free combination of the $^{35}$S and $^{14}$C β-spectra yielded reduced $\chi^2$ value of 1.06. A fraction of decays of $^{14}$C derived from the β-spectrum fit, 1.4 ± 0.1 % of the total decay rate, agreed with a value of 1.34% measured for $^{14}$C alone during the $^{35}$S + $^{14}$C source preparation. The spectra of residuals of the composite β-spectrum fit are shown in Fig. (**29**). Further investigation of the $^{35}$S β-spectrum [195] clarified that the 17 keV neutrino claim, reported in previous studies of this isotope [196], [192] originated mostly from an improper treatment of the electron energy losses in the source itself. Still another exploration of $^{35}$S β-spectrum revealed that a thin metallic diaphragm in front a Si(Li) detector can generate a false 17 keV neutrino admixture of 0.3% [197].

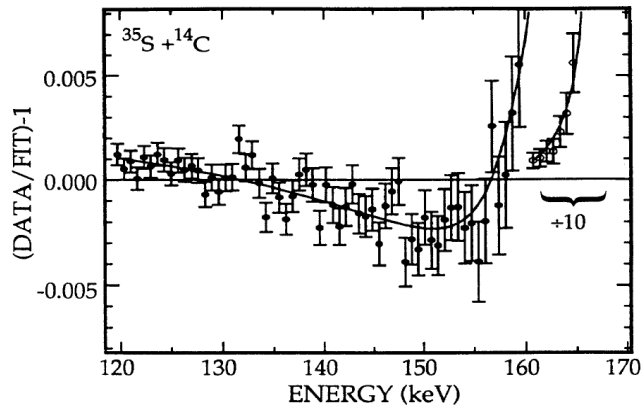

**Fig. (29).** Residuals from fitting the $^{35}$S β-spectrum shape into a measured β-spectrum of the composite ($^{35}$S + $^{14}$C) source [194]. The solid curve indicates residuals expected from the known admixture of $^{14}$C.

Although now disproved, the claim for the 17 keV neutrino (in a similar manner to the 30 eV claim discussed in Sect. 3.3) pushed forward the development of precision β-ray spectroscopy. The relevant experiments with both positive and negative results are listed in refs. [198], [199]. Their detailed account can be found in refs. [200], [201]. "The rise and fall of the 17 keV neutrino" was also briefly described in ref. [202].

Nevertheless, searches for kinks in β-ray spectra in a wider region of $m_4$ continued. Hiddemann *et al.* [203] searched for a signature of the sterile neutrinos in their β-spectrum of tritium taken with a magnetic spectrometer but no evidence was found in the mass region 10 eV $\leq m_4 \leq$ 4 keV. Holzschuh *et al.* [204] explored the β-spectrum of $^{63}$Ni with a magnetic spectrometer [111] and found the $m_4$ mixing probability to be less than $1 \cdot 10^{-3}$ at the 95% CL for neutrino masses in the range from 4 to 30 keV. In their study of the β-spectrum of $^{35}$S, the same upper limit of $|U_{e4}|^2$ was determined for $10 \leq m_4 \leq 90$ keV [205]. Galeazzi *et al.* [206] searched for the kinks in the low-energy β-spectrum of $^{187}$Re ($Q_\beta$ = 2.44 keV, $T_{1/2}$ = 4.3·10$^{10}$y) with cryogenic microcalorimeters, mentioned above in Sect. 3.5.2. A single crystal of superconducting rhenium with a mass of 1.5 mg and activity of about 1.1 Bq served as a source as well as an absorber of $^{187}$Re β-rays. No evidence for an additional neutrino mass state $m_4$ was observed. The upper limits of an $|U_{e4}|^2$ admixture between 0.9% and 4.4% for $m_4$ = 1000 and 200 eV, respectively, were established at the 95% CL.

An agreement between measured and theoretical β-spectra is often reached only after adding one or two shape correction parameters without clear physical meaning. Such a procedure not only deteriorates the sensitivity of the experiment (since it increases the number of fitted parameters) but it can also hide some spectrum details or create artificial ones[22]. For example, Apalikov *et al.* [208] investigated the β-spectrum of $^{35}$S with a magnetic spectrometer [109] and found limits for $5 \leq m_4 \leq 80$ keV. An admixture of the 17.1 keV neutrino was set to be less than 0.17% at the 90% CL. However, in order to reproduce their measured β-spectrum shape, the authors had to introduce a phenomenological correction in the form

$$f_{\text{corr}}(E) = [1 + \alpha(E_0 - E)] + [1 + \alpha'(E_0 - E)^2] \qquad (18)$$

Thus the number of fitted parameters increased by two. The $f_{\text{corr}}(E)$ varied from 1.012 to 1.29 for $m_4$ between 5 and 80 keV, respectively. For $m_4$ = 17.1 keV the spectrum correction amounted to 4.5%.

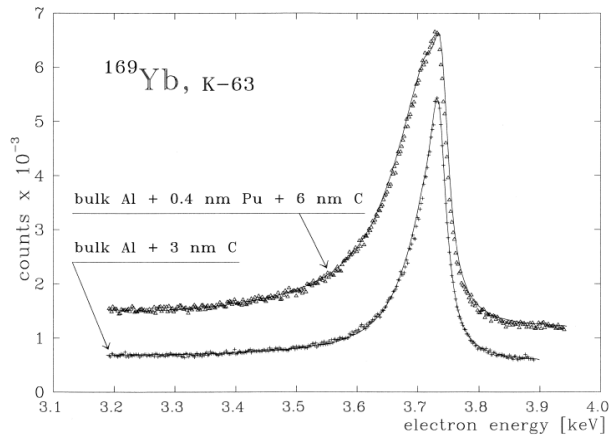

**Fig. (30).** The K-shell internal-conversion electron line of the 63.1 keV transition in $^{169}_{69}Tm$ measured with an original and modified $^{169}$Yb source [209]. The original source was prepared by vacuum evaporation onto Al backing and was contaminated by a usual hydrocarbon over layer[23] (described by a 3 nm thick carbon layer in Monte Carlo calculations). In the case of the modified source a supplementary Pu layer of average thickness of 0.4 nm was evaporated where electrons of the kinetic energy of 3.7 keV suffered additional elastic and inelastic scattering. The agreement of measured and calculated (smooth curves) shapes indicates correctness of the electron loss function in plutonium applied in calculations of the $^{241}$Pu β-spectrum shape. The natural width of this conversion line is 32.0 eV [212].

---

[22] The observability of the 17 keV neutrino in a β-spectrum distorted by unknown effects was examined by statistical methods in ref. [207].
[23] Such a contamination is known to exist on the surface of all sources used in the electron spectrometry of radionuclides except for gaseous ones and those prepared in situ under ultra-high vacuum conditions.

Within the search for an additional neutrino mass-state in the β-decay of $^{241}$Pu ($Q_\beta$ = 20.8 keV, $T_{1/2}$ = 14.4 y) [209], the electron scattering and energy losses within a vacuum-evaporated source were carefully examined. The measurements were carried out with two electrostatic spectrometers [210], [211]. The individual elastic and inelastic scattering events were simulated using the Monte Carlo method. The quality of the approximation was tested by measuring the shape of the 3.7 keV conversion electron line of the $^{169}$Yb source, see Fig. (30). This approach allowed the description of the measured part of the $^{241}$Pu spectrum (from 0.2 to 9.2 keV) down to 2 keV without any artificial fitting parameter.

Fig. (31) demonstrates why the β-spectra intended for sterile neutrino searches should be measured with sufficient instrumental resolution. Lines observed on both continuous β-spectra of ground-state-to-ground-state decays of $^{241}$Pu and $^{63}$Ni at energies of about 260 and 510 eV are the KLL Auger lines of carbon and oxygen. The K-shell vacancies were created by the impact of β-particles from the sources on atoms in the contamination layer. The observed line intensity is in good agreement with the prediction based on the cross-section for electron-impact ionization amounting $(10^4 – 10^5) \cdot 10^{-24}$ cm$^2$ for the β-particles of given energy [213] and activity of the sources (6.5 MBq for $^{241}$Pu and 1 MBq for $^{63}$Ni). If the spectra were recorded at a worse instrumental resolution, the Auger lines would smear out and one could erroneously interpret the observed deviation as an anomaly in the β-spectrum shape.

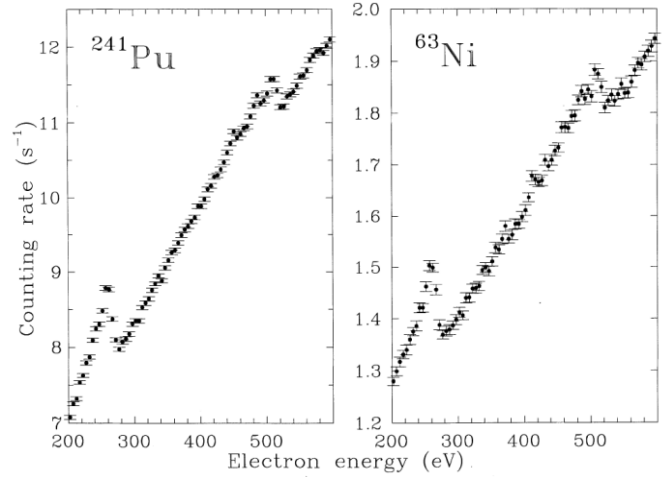

**Fig. (31).** The KLL-Auger lines of carbon and oxygen superimposed on continuous β-spectra of $^{241}$Pu and $^{63}$Ni [214]. The spectra of the vacuum-evaporated sources were recorded with an electrostatic spectrometer [210], [215] by scanning the shown energy regions 10 000 and 18 000 times[24]. The measurement time was 1 s per point in every scan.

Recently, Troitsk physicists extended the analysis of tritium β-spectra gathered in their search for the effective mass $m_{\nu_e}$ of light active neutrinos [39] and using the maximum-likelihood method they looked for any signatures of sterile neutrinos. A possible admixture of the neutrino mass state with $20 \leq m_4 \leq 100$ eV was found to be less than 1 % at the 95 % CL [217], [218]. The surprising feature of their analysis is that this upper limit did not depend on systematic uncertainties of their experiment. This was not the case in their analysis of the same β-spectra for $m_{\nu_e}$ where statistical and systematic uncertainties played about the same role [39]. Even a new analysis of tritium β-spectra of the Mainz neutrino mass experiment [38], involving detailed examination of systematic uncertainties, did not reveal any admixture of the fourth neutrino mass state [219].

A broad energy range of β-spectroscopic searches for sterile neutrinos is illustrated in Fig. (32). Relevant data is summarized in the section "Kink search in nuclear β-decay" of the current issue of the Review of particle physics [8].

---

[24] Compatibility of parts of β-spectra collected during various time intervals can be tested e.g. by sensitive statistical tests [216].

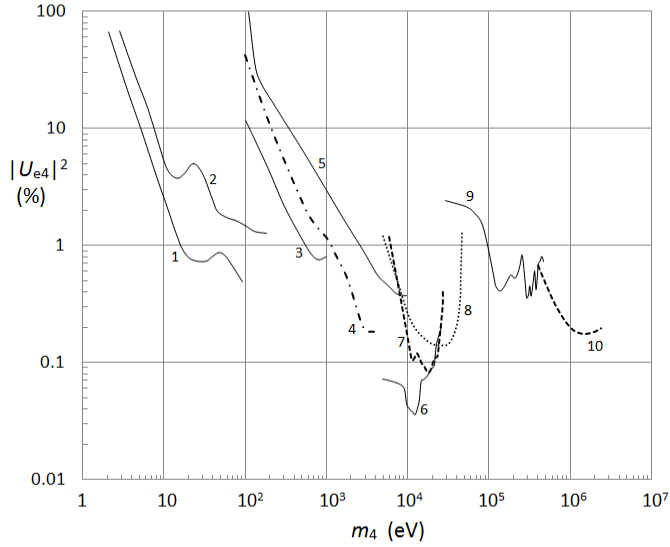

**Fig. (32).** Best upper limits on the admixture $|U_{e4}|^2$ of sterile neutrinos derived from measured β-ray spectra. All upper limits refer to the 95 % CL except those for $^{64}$Cu [186] and $^{3}$H [219] that were published at the 90% CL. The numbers denote the following radionuclides and their investigators: 1 – $^{3}$H [218], 2 – $^{3}$H [219], 3 – $^{187}$Re [206], 4 – $^{3}$H [203], 5 – $^{3}$H [114], 6 – $^{63}$Ni [204], 7 – $^{63}$Ni [193], 8 – $^{35}$S [194], 9 – $^{64}$Cu [186], 10 – $^{20}$F [189].

### 3.7.2 Future experiments

Formaggio and Barret [220] examined whether the KATRIN experiment [40] with intended sensitivity of 200 meV for light active neutrinos could verify the claim of Mention *et al.* for the possible existence of a sterile neutrino with $|\Delta m_{41}^2| = |m_4^2 - m_1^2| > 1.5$ eV$^2$ and $\sin^2(2\theta_4) = 0.14 \pm 0.08$ at 95% CL [221][25]. The authors [220] estimated that KATRIN after three years of data-taking should be able to place the limits $\sin^2(2\theta_4) \leq 0.08$ and 0.20 for $|\Delta m_{41}^2| > 4$ and 2 eV$^2$, respectively. The two values of $\sin^2(2\theta_4)$ correspond to $|U_{e4}|^2 = 0.02$ and 0.05. Similar results were obtained in ref. [224] with the conclusion that, for mixing angles $\sin^2(2\theta_4) \leq 0.1$, the KATRIN experiment could provide the most stringent limit on the active-sterile mass square difference $|\Delta m_{41}^2|$.

Further possibilities of KATRIN-like setups were examined by Sejersen Riis and Hannestad [225] who considered the 3+2 scenario, i.e. the mixing of three almost massless active neutrinos with two additional massive sterile neutrinos coupled to $\bar{\nu}_e$. Their Monte-Carlo-generated tritium β-spectra included the sterile mass states $m_4$ and $m_5$ up to 6.4 eV and mixings $|U_{e4}|^2$ and $|U_{e5}|^2$ in the interval from $5.5 \cdot 10^{-4}$ to 0.18. Projected parameters of the KATRIN setup, including an energy resolution of 0.93 eV, a background of 0.01s$^{-1}$ and systematic uncertainty of 0.017 eV$^2$ [142], were assumed. Possible future improvements such as a higher β-spectrum amplitude, better resolution and lower background were probed as well. The authors concluded that the KATRIN experiment should be sensitive enough to confirm or disprove sterile neutrinos with $\Delta m_{41}^2 = 6.49$ eV$^2$, $|U_{e4}|^2 = 0.12$ and $|U_{e5}|^2 = 0.11$, as suggested by some cosmologists. In the $2 \times 2$ mixing scheme (one active light neutrino with the effective mass $m_{\nu_e}$ according to Eq. 5 and one sterile neutrino), any of the mass states $m_4$ with $|U_{e4}|^2 \geq 0.055$ should be detected with 3σ confidence. The same sensitivity is expected for any of the mixing angles for the mass state with $m_4 \geq 3.2$ eV.

Mertens *et al.* [82] recently scrutinized the sensitivity of future tritium β-decay experiments towards hypothetical keV-scale sterile neutrinos that are one of the candidates for dark matter of the universe, see e.g. [10]. Experimental considerations of the authors were based on the statistics of $10^{18}$ β-decay events occurring in the KATRIN gaseous tritium source during the three-year exposure time. Three possible measurement modes were taken into account: (i) a point-by-point measurement, with variable retardation of the electrostatic spectrometer yielding the β-spectrum in its integral form, (ii) an application of the KATRIN high-luminosity tritium source together with a future semiconductor spectrometer consisting of up to $10^6$ independent pixels and capable of operating at $10^5$ events per second, (iii) the KATRIN tritium source and electrostatic spectrometer but operating in the time-of-flight regime [226] where the last two modes yield the β-spectra in their differential form. As can be seen in Fig. (33), the differential spectra exhibit an extremely high sensitivity to $|U_{e4}|^2$ even in the case of a moderate spectrometer resolution. Note, however, that the limits shown comprise only the statistical sensitivity; the systematic uncertainties are not included.

---

[25] The original version of the paper [221] involved a value of $\sin^2(2\theta_{sterile}) = 0.17 \pm 0.08$ at 95% CL that was utilized in the analysis of Formaggio and Barret [220]. Later, the reported reactor anomaly of 2.5 σ in the short baseline neutrino oscillations [221] was weakened to 1.4 σ when the recently measured value of the mixing angle $\theta_{13}$ was taken into account [222]; the former ratio of observed to predicted neutrino flux of 0.943 ± 0.023 was thus reduced to 0.959 ± 0.009 (experimental uncertainty) ± 0.027 (flux systematics). Garvey *et al.* [223] took into account that up to 30% of β-decays of all fission products may be of the first forbidden type whose properties are not adequately known. This fact modifies the energy spectrum of antineutrinos emitted by nuclear reactor. From the authors' analysis it follows that the uncertainty on the neutrino flux should be expanded beyond 4%.

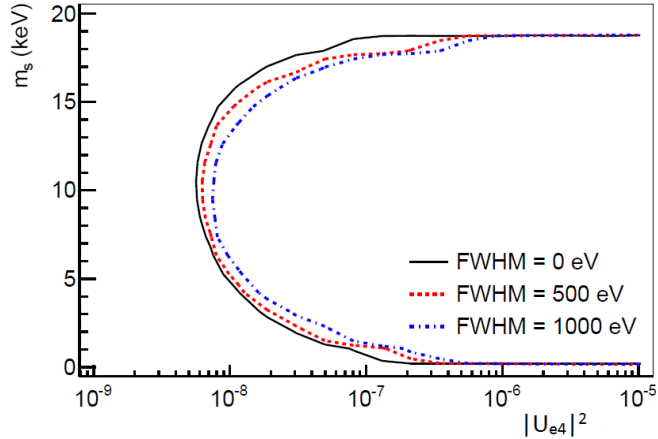

**Fig. (33).** The 90% exclusion limit for an admixture of keV sterile neutrinos obtained from simulated β-spectra [82] that correspond to differential measurements of the KATRIN-like tritium source with a future multi-pixel semiconductor spectrometer of indicated energy resolution.

The authors [227] estimated quantitatively the influence of eight different uncertainties of the shape of the theoretical β-spectrum, such as decay into excited final states, nuclear correction terms, *etc*, and concluded that these uncertainties decrease the sensitivity to keV neutrinos by about one order of magnitude. Ensuring the relative standard deviation of individual points of the overall β-spectrum on the level of $10^{-6}$ during 3 measurement years is an extremely challenging task. Therefore the authors considered an alternative approach where the keV sterile neutrinos would be searched in a β-spectrum with the same tritium source as in the KATRIN experiment but with a high-resolution detector. Their simulations showed that the wavelet transform, i.e. the convolution of the β-spectrum with specific window functions, is largely insensitive to smooth spectral deformations that are caused e.g. by slow changes of the measurement conditions. A statistical sensitivity to sterile-to-active neutrino mixing $|U_{e4}|^2 = 10^{-6}$ is expected for the detector resolution of FWHM ≈ 100 eV. Nevertheless, the most reliable way to correct experimental results is a sufficiently precise knowledge of all systematic effects. Such approach may require a series of side measurements as exemplified by the Mainz neutrino mass experiment [38].

Filianin *et al.* [228] are developing methods of cryogenic microcalorimetry and high-precision Penning-trap mass spectrometry in order to probe the possible existence of sterile neutrinos with masses from 0.5 to 100 keV. The authors considered the electron-capture decay of eight isotopes including $^{163}$Ho, $^{179}$Ta, $^{193}$Pt and $^{235}$Np. They also examined a rather complex contribution of a sterile neutrino within the shape of an electron-capture spectrum; see Sect. 3.5.2 and refs. [150], [152].

Abdurashitov *et al*. [229] proposed a measurement of the tritium β-spectrum with proportional counters that has already proven long time stability in the SAGE experiment on solar neutrinos. The authors estimated that an analysis of a β-spectrum measured with relatively poor resolution of 10–15% and containing a total of $10^{12}$ registered β-particles, could reach a sensitivity of $10^{-3}$ to $10^{-5}$ for an admixture of sterile neutrinos in the mass range from 1 to 8 keV. This estimate took into account the deterioration of the shape of the spectrum due to β-decays near the counter inner wall as well as a pile-up effect at counting rates up to $10^6$ s$^{-1}$. However, one should also include the β-decay of tritium adsorbed on the counter walls – an effect that has already been carefully examined and proved to be substantial in ref. [230] in connection with the 17 keV neutrino claim.

The Troitsk experimental set-up for an investigation of tritium β-spectra was recently equipped with a new MAC-E-filter of twice the diameter and ten times the volume of the older one [39], [132]. The field of the main magnet was increased by 20%. The energy resolution at 18.6 keV is now 1.5 eV (FW), more than twice better than previously. The new facility should enable searching for sterile neutrinos in a few keV mass range with a sensitivity for $|U_{e4}|^2$ up to ~$2·10^{-4}$ and a further improvement of an order of magnitude after future upgrades [231].

The role of sterile neutrino warm dark matter[26] was examined using $^{187}$Re and $^3$H β-decays in ref. [233]. Theoretical aspects of sterile neutrinos and right-handed currents were treated in the frame of the anticipated KATRIN experiment in ref. [234]. A possibility to record signatures of extra dimensional sterile neutrinos in this experiment was examined in ref. [235].

## 4. CONCLUSION

It has been a hundred years since James Chadwick convincingly proved that the energy spectrum of β-rays is continuous[27]. This surprising finding, together with an unexpected spin of several nuclei like $^{14}_{7}N$, initiated a novel concept of the neutrino by Wolfgang Pauli [1] and of the weak interactions by Enrico Fermi [2]. In 1948 β-ray spectroscopy provided the first upper limit of the neutrino rest mass [92]. The next two generations of β-ray spectroscopists succeeded in improving this limit by three orders of magnitude to the present value of the effective electron neutrino mass $m_{\nu_e} \leq 2$ eV [8]. While the oscillation experiments provide the differences $m_i^2 - m_j^2$ with increasing precision, the values of $m_i$ remain unknown. The upcoming KATRIN experiment with its projected sensitivity of 200 meV [40] should explore an almost full quasi-degenerate region of neutrino masses, see Fig. (**1**). It is hoped that new β-spectroscopic approaches, including those listed in this review, will lead to an even higher sensitivity.

The higher sensitivity, allowing studies of the inverted neutrino mass hierarchy, is expected also from projected 0νββ experiments, see e.g. [48]. Of course, the 0νββ process may not exist or, for a particular value of Majorana phases, the effective neutrino mass $m_{\beta\beta}$ may be close to zero, even for $m_i \neq 0$. The region of normal mass hierarchy seems to be now accessible only via future cosmological observations. Regardless of remarkable progress, their interpretation is yet strongly model dependent and therefore may suffer from large systematic uncertainties.

The analysis of SBL neutrino oscillation experiments indicates the existence of sterile neutrinos in the eV mass range. Further tests are listed in ref. [9]. Simulations predict that the KATRIN β-ray spectrometer will have a high sensitivity also for these hypothetical particles [220], [224], [225].

Due to the importance of the absolute value of the neutrino mass for particle physics and cosmology, its model independent determination remains a great challenge not only for β-ray spectroscopists.

**Conflict of interest**
The authors confirm that this article content has no conflict of interest.

**Acknowledgement**
This work was supported by the GACR (contract P203/12/1896) and the ASCR (contract IRP AV0Z10480505).

**REFERENCES**
[1]   Pauli W. The letter to Meitner L *et al*. In: Kroning R, Weisskopf VF, Eds. Collected scientific papers by Wolfgang Pauli. Wiley Interscience, New York, 1964: Vol. 2, p. 1336.
[2]   Fermi E. Versuch einer Theorie der β-Strahlen, Z Phys 1934; 88: 161-77. English translation in: Wilson FL. Fermi's theory of beta decay. Amer J Phys 1968; 36: 1150-60.
[3]   Reines F, Cowan CL, Harrison FB, McGuire AD, Kruse HW. Detection of the free antineutrino. Phys Rev 1960; 117: 159-73.
[4]   Danby G, Gaillard J-M, Goulianos K, *et al*. Observation of high-energy neutrino reactions and the existence of two kinds of neutrinos. Phys Rev Lett 1962; 9: 36-44.
[5]   Kodama K, Ushida N, Andreopoulos C, *et al*. Observation of tau neutrino interactions. Phys Lett B 2011; 504: 218-24.
[6]   Shrock RE. New tests for and bounds on neutrino masses and lepton mixing. Phys Lett B 1980; 96: 159-64.
[7]   Boehm F, Vogel P. Physics of Massive Neutrinos. Cambridge University Press, Cambridge, 2nd edition, 1992.

---

[26] According to ref. [232] these sterile neutrinos could not only be warm but also cold dark matter.

[27] It was achieved by means of a magnetic spectrometer equipped with a Geiger counter. Former recording of electron intensity with photographic plates enabled rather precise measurement of the energy of monoenergetic internal conversion electrons but due to nonlinearity of the plate blackening did not allow reliable investigation of β-rays.

[8]     Olive KA, Agashe K, Amsler C, *et al*. Review of particle properties. Chin Phys C 2014; 38: 090001, 1-1676.
[9]     Gariazzo S, Giunti C, Laveder M, Li YF, Zavanin EM. Light sterile neutrinos. arXiv: 1507.08204 (40pp).
[10]    Abazajian KN, Acero MA, Agarwalla SK, *et al*. Light sterile neutrinos: a white paper. arXiv: 1204.5379 (269pp).
[11]    Vogel P, Wenn LJ, Zhang C. Neutrino oscillation studies with reactors. Nature Comm 2015; 6: 6935, 1-12. arXiv: 1503.01059.
[12]    Fukuda Y, Hayakawa T, Ishihava E, *et al*. Evidence for oscillation of atmospheric neutrinos. Phys Rev Lett 1998; 81: 1562-7.
[13]    Abazajian KN, Akiri T, Albright C. Neutrinos. arXiv: 1310.4340 (89pp).
[14]    Barker GJ, Biassoni M, De Rujula A, *et al*. The future of neutrino mass measurements: terrestrial, astrophysical and cosmological measurements in the next decade. arXiv: 1309.7810 (63pp).
[15]    Petcov ST. The nature of massive neutrinos. Adv High Energy Phys 2013; 2013: 852987, 1-20. arXiv: 1303.5819.
[16]    McKeown RD. Neutrinos in nuclear physics. arXiv: 1412.1364 (8pp).
[17]    Vissani F. Neutrino sources and properties. arXiv: 1412.8386 (15pp).
[18]    Goodman M. Neutrino physics in 2020. arXiv: 1402.6207 (5pp).
[19]    Giunti C, Kim ChW. Fundamentals of Neutrino Physics and Astrophysics. Oxford University Press: Oxford; 2011.
[20]    Zuber K. Neutrino Physics. 2nd ed. Taylor & Francis: New York; 2011.
[21]    Klapdor-Kleingrothaus HV, editor. Seventy Years of Double Beta Decay. From Nuclear Physics to Beyond-Standard-Model Particle Physics. World Scientific: Singapore; 2010.
[22]    Lesgourgues J, Mangano G, Miele G, Pastor S. Neutrino Cosmology. University Press: Cambridge; 2013.
[23]    Barner V, Marfatia D, Whisnant K. The physics of neutrinos. Princeton Uni Press: Princeton; 2012.
[24]    Smirnov AYu. Riddle of the neutrino mass. Nucl Phys B Proc Suppl 2015 (in press). arXiv: 1502.04530 (6pp).
[25]    Kirilova D. Neutrinos from the early universe and physics beyond standard models. Open Phys 2015; 13: 22-33. arXiv: 1407.1784.
[26]    Volpe C. Recent advances in neutrino astrophysics. arXiv: 1411.6533 (6pp).
[27]    Haxton WC. Neutrino astrophysics. In: Stock R, Ed. Encyclopedia of nuclear physics and its applications, Wiley, New York, 2013; pp. 353-94. arXiv: 1209.3743.
[28]    Bilenky SM. Neutrino oscillations: brief history and present status. arXiv: 1408.2864 (42pp).
[29]    Bilenky SM, Giunti C. Neutrinoless double-beta decay: a probe of physics beyond the standard model. arXiv: 1411.4791 (58pp).
[30]    Gómez-Cadenas JJ, Martín-Albo J. Phenomenology of neutrinoless double beta decay. arXiv: 1402.00581 (35pp).
[31]    Drexlin G, Hannen V, Mertens S, Weinheimer C. Current direct neutrino mass experiments. Adv High Energy Phys 2013; 293986, 1-39.
[32]    Archidiacono M, Fornengo N, Gariazzo S, Giunti C, Hannestad S, Laveder M. Light sterile neutrinos after BICEP-2. J Cosm Astropart Phys 2014; 06: 031, 1-11. arXiv: 1404.1794.
[33]    Robertson RGH, Knapp DA. Direct measurements of neutrino mass. Ann Rev Nucl Part Sci 1988; 38: 185-215.
[34]    Holzschuh E. Measurement of the neutrino mass from tritium β-decay. Rep Prog Phys 1992; 55: 1035-91.
[35]    Jaros JA. Searches for massive neutrinos in nuclear beta decay. Int J Mod Phys A, Proc Suppl 1993; 3B: 149-71.
[36]    Otten EW, Weinheimer C. Neutrino mass limit from tritium β decay. Rep Prog Phys 2008; 71: 086201, 1-89. arXiv: 0909.2104.
[37]    Klapdor-Kleingrothaus HV, Krivosheina IV. The evidence for the observation of 0νββ decay: The identification of 0νββ events from the full spectra. Mod Phys Lett A 2006; 21:1547-66.
[38]    Kraus C, Bornschein B, Bornschein L *et al*. Final results from phase II of the Mainz neutrino mass search in tritium β decay. Eur Phys J C 2005; 4: 447-68.
[39]    Aseev VN, Belesev AI, Berlev AI *et al*. An upper limit on electron antineutrino mass from Troitsk experiment. Phys Rev D 2011; 84: 112003, 1-9. arXiv: 1108.5034.
[40]    http://www.katrin.kit.edu/.
[41]    Vogel P, Piepke A. Neutrinoless double-β decay. In the Review of particle properties. Chin Phys C 2014; 38: 090001, 1-1676.
[42]    Vergados JD, Ejiri H, Šimkovic F. Theory of neutrinoless double beta decay. Rep Prog Phys 2012; 75:106301, 1-52. arXiv: 1205.0649.
[43]    Duerr M, Lindner M, Merle A. On the quantitative impact of the Schechter-Valle theorem. J High Ener Phys 2011; 06: 091, 1-18.
[44]    Engel J, Šimkovic F, Vogel P. Chiral two-body currents and neutrinoless double-beta decay in the QRPA. Phys Rev C 2014; 89: 064308, 1-5. arXiv: 1403.7260.
[45]    Moe M, Vogel P. Double beta decay. Ann Rev Nucl Part Sci 1994; 44: 247-83.
[46]    Barabash AS. Double beta decay: historical review of 75 years of research. Phys Atom Nucl 2011; 74: 603-13. arXiv: 1104.2714.
[47]    Avignone III FT, Elliot SR, Engel J. Double beta decay, Majorana neutrinos, and neutrino mass. Rev Mod Phys 2008; 80: 481-516.
[48]    Schwingenheuer B. Status and prospects of searches for neutrinoless double beta decay. Ann Phys (Berlin) 2013; 525: 269-80. arXiv:1210.7432.
[49]    Dell'Oro S, Marcocci S, Vissani F. New expectations and uncertainties on neutrinoless double beta decay. Phys Rev D 2014; 90: 033005, 1-8. arXiv: 1404.2616.
[50]    Garfagnini A. Neutrinoless Double Beta Decay Experiments. arXiv: 1408.2455 (10pp).
[51]    Tosi D. The search for neutrino-less double-beta decay. arXiv: 1402.1170 (11pp).
[52]    Tornow W. Search for neutrinoless double-beta decay. arXiv: 1412.0734 (10pp).
[53]    Klapdor-Kleingrothaus HV, Dietz A, Harney HL, Krivosheina IV, *et al*. Evidence for neutrinoless double beta decay. Mod Phys Lett A 2001; 16: 2409-20.
[54]    Aalseth CE, Avignone III FT, Barabash A, *et al*. Comment on "Evidence for neutrinoless double beta decay". Mod Phys Lett A 2002; 17: 1475-8.
[55]    Feruglio F, Strumia A, Vissani F. Neutrino oscillations and signals in β and 0νββ experiments. Nucl Phys B 2002; 637: 345-77.
[56]    Ackerman K-H, Agostini M, Allardt M *et al*. The GERDA experiment for the search of 0νββ decay of $^{76}$Ge. Eur Phys J C 2013; 73: 2330, 1-29. arXiv: 1212.4067.
[57]    Agostini M, Allardt M, Andreotti E, *et al*. Results on neutrinoless double beta decay of $^{76}$Ge from Gerda Phase I. Phys Rev Lett 2013; 111: 122503, 1-6. arXiv: 1307.4720.

[58]   Arnold R, Augier C, Baker JD *et al.* Search for neutrinoless double-beta decay of $^{100}$Mo with the NEMO-3 detector. archiv: 1311.5695 (5pp).
[59]   Andreotti E, Arnaboldi C, Avignone III FT, *et al.* $^{130}$Te neutrinoless double-beta decay with CUORICINO. Astropart Phys 2011; 34: 822-831.
[60]   Albert JB, Auty DJ, Barbeau PS, *et al.* Search for Majorana neutrinos with the first two years of EXO-200 data. Nature 2014; 510: 229-34. arXiv: 1402.6956.
[61]   Gando A, Gando, Hanakago H, *at al.* Limit on neutrinoless ββ decay of $^{136}$Xe from the first phase of KamLAND-Zen and comparison with the positive claim in $^{76}$Ge. Phys Rev Lett 2013; 110: 062502, 1-5. arXiv:1211.3863.
[62]   Auger M, Auty DJ, Barbeau PS. Search for neutrinoless double-beta decay in $^{136}$Xe with EXO-200. Phys Rev Lett 2012; 109: 032505, 1-6. arXiv: 1205.5608.
[63]   Eisenstein DJ. Neutrinos and large-scale structure. AIP Conf Proc 2015; 1666: 14002, 1-6.
[64]   Ibarra A. Neutrinos and dark matter. AIP Conf Proc 2015; 1666: 14004, 1-8.
[65]   Lattanzi M, Lineros RA, Taoso M. Connecting neutrino physics with dark matter. New J Phys 2014; 16: 125012, 1-20. arXiv: 1406.0004.
[66]   Ade PAR, Aghanim N, Armitage-Caplan NC, *et al.* Planck 2013 results XVI. Cosmological parameters. Astron Astrophys 2014; 571: A16, 1-66. arXiv: 1303.5076.
[67]   Beutler F, Saito S, Brownstein JR, *et al.* The clustering of galaxies in the SDSS-III baryon oscillation spectroscopic survey: signs of neutrino mass in current cosmological datasets. Mon Not Roy Astron Soc 2014; 444: 3501-16. arXiv: 1403.4599.
[68]   Beringer J, Arguin J-F, Barnett RM, *et al.* Review of particle properties. Phys Rev D 2012; 86: 010001 (1525pp).
[69]   Palanque-Delabrouille N, Yèche C, Lesgourgues J, *et al*. Constraint on neutrino masses from SDSS-III/BOSS Lyα forest and other cosmological probes. J Cosm Astropart Phys 2015; 02/045, 1-38. arXiv: 1410.7244.
[70]   Costanzi M, Sartoris B, Viel M, Borgani S. Neutrino constraints: what large-scale structure and CMB data are telling us? arXiv: 1407.8338 (22pp).
[71]   Zablocki A. Constraining neutrinos and dark energy with galaxy clustering in the dark energy survey. arXiv: 1411.7387 (27pp).
[72]   Vissani F. Non-oscillation searches of neutrino mass in the age of oscillations. Nucl Phys B Proc Suppl 2001; 100: 273-5.
[73]   Farzan Y, Peres OLG, Smirnov A.Yu. Neutrino mass spectrum and future beta decay experiments. Nucl Phys B 2001; 612: 59-97.
[74]   Farzan Y, Smirnov AYu. On the effective mass of the electron neutrino in beta decay. Phys Lett B 2003; 557: 224-32.
[75]   Assamagan K, Brönnimann Ch, Daum M, Frosch R, Kettle P-R, Wigger C. Search for a heavy neutrino state in the decay $\pi^+ \to \mu^+ + \nu_\mu$. Phys Lett B 1998; 434: 158-62.
[76]   Barate R, Buskulic D, Decamp D, *et al.* An upper limit on the τ neutrino mass from three- and five-prong tau decays. Eur Phys J C 1998; 2: 395-406.
[77]   Loredo TJ, Lamb DQ. Bayesian analysis of neutrinos observed from supernova SN 1987A. Phys Rev D 2002; 65: 063002, 1-39. arXiv: 0107260.
[78]   Vissani F. Comparative analysis of SN1987A antineutrino fluency. J Phys G 2015; 42: 013001, 1-39. arXiv: 1409.4710.
[79]   Cirelli M, Marandela G, Sturmia A, Vissani F. Probing oscillations into sterile neutrinos with cosmology, astrophysics and experiments. Nucl. Phys. B 2005; 708: 215-67.
[80]   Del'Oro S, Marcocci S, Viel M, Vissani F. The contribution of light Majorana neutrinos to neutrinoless double beta decay and cosmology. arXiv: 1505.02722 (4pp).
[81]   Lesgourgues J, Pastor S. Neutrino cosmology and PLANCK. New J Phys 2014; 16: 065002, 1-24. arXiv: 1404.1740.
[82]   Mertens S, Lasserre T, Groh S, *et al*. Sensitivity of next-generation tritium beta-decay experiments for keV-sterile neutrinos. J Cosm Astropart Phys 2014 (in press). arXiv: 1409.0920 (23pp).
[83]   Bergkvist K-E. A high-luminosity, high-resolution study of the end-point behavior of the tritium β-spectrum (I). Basic experimental procedure and analysis with regard to neutrino mass and neutrino degeneracy. Nucl Phys B 1972; 39: 317-70.
[84]   Williams RD, Koonin SE. Atomic final-state interactions in tritium decay. Phys Lett C 1983; 27: 1815-7.
[85]   Bodine LI, Parno DS, Robertson RGH. Assessment of molecular effects on neutrino mass measurements from tritium beta decay. Phys Rev C 2015; 91: 035505, 1-24. arXiv: 1502.03497.
[86]   Doss N, Tennyson J. Excitations to the electronic continuum of $^3$HeT$^+$ in investigations of T$_2$ β-decay experiments. Europhys News 2008; 39: No 5, p 16.
[87]   Doss N, Tennyson J, Saenz A, Jonsell S. Molecular effects in investigations of tritium molecule *β* decay endpoint experiments. Phys Rev C 2006; 73: 025502, 1-10.
[88]   Doss N, Tennyson J. Excitations to the electronic continuum of $^3$HeT$^+$ in investigations of T$_2$ β-decay experiments. J Phys B: At Mol Opt Phys 2008; 41: 125701, 1-10.
[89]   Kašpar J, Ryšavý M, Špalek A, Dragoun O. Effect of energy scale imperfections on results of neutrino mass measurements from β-decay. Nucl Instr Meth Phys Res A 2004; 527: 423-431.
[90]   Otten EW, Bonn J, Weinheimer Ch. The *Q*-value of tritium β-decay and the neutrino mass. Int J Mass Spect 2006; 251: 173-8.
[91]   Myers EG, Wagner A, Kracke H, Wesson BA. Atomic masses of tritium and helium-3. Phys Rev Lett 2015; 115: 013003, 1-5.
[92]   Cook CS, Langer M, Price Jr HC. The β-spectrum of $^{35}$S and the mass of the neutrino. Phys Rev 1948; 73: 1395.
[93]   Langer LM, Cook CS. A high resolution nuclear spectrometer. Rev Sci Inst 1948; 19: 257-62.
[94]   Curran SC, Angus J, Cockroft AL. II. Investigation of soft radiations by proportional counters. Phil Mag Series 7, 1949; 40: 36-52.
[95]   Hanna GC, Kirkwood DHW, Pontecorvo B. High multiplication proportional counters for energy measurements. Phys Rev 1949; 75: 985-6.
[96]   Hanna GC, Pontecorvo B. The β-spectrum of $^3$H. Phys Rev 1949; 75: 983-4.
[97]   Curran SC, Angus J, Cockroft AL. The beta-spectrum of tritium. Phys Rev 1949; 76: 853-4.
[98]   Langer LM, Moffat RJD. The beta-spectrum of tritium and the the mass of the neutrino. Phys Rev 1952; 88: 689-94.
[99]   Hamilton DR, Alford WA, Gross L. Upper limits on the neutrino mass from the tritium beta spectrum. Phys Rev 1953; 92: 1521-5.
[100]  Hamilton DR, Gross L. An electrostatic beta-spectrograph. Rev Sci Instr 1950; 21: 912-7.

[101]    Salgo RC, Staub HH. Re-determination of the β-energy of tritium and its relation to the neutrino rest mass and the Gamow-Teller matrix element. Nucl Phys A 1969; 138: 417-28.
[102]    Daris R, St.-Pierre C. Beta decay of tritium. Nucl Phys A 1969; 138: 545-55.
[103]    Bergkvist K-E. Combined electrostatic-magnetic β-spectrometer with high luminosity. Part I: Principles and basic properties. Ark Fys 1964; 27: 383-437.
[104]    Bergkvist K-E. Combined electrostatic-magnetic β-spectrometer with high luminosity. Patr II: Application of the methods to an $r_0 = 50$ cm iron yoke double focusing spectrometer. Ark Fys 1964; 27: 439-82.
[105]    Bergkvist K-E. A high-luminosity, high-resolution study of the end-point behavior of the tritium β-spectrum (II). The endpoint energy of the spectrum. Comparison of the experimental axial-vector matrix element with predictions based on PCAC. Nucl Phys B 1972; 39: 371-406.
[106]    Röde B, Daniel H. Measurement of the $^3$H β-ray spectrum and determination of an upper limit for the electron-antineutrino rest mass. Lett Nuo Cim 1972; 5: 139-43.
[107]    Röde B. Messung des $^3$H-beta spectrums und Bestimmung der Obergrenze für die Ruhemasse des elektronischen antineutrinos. Z Naturforsch 74; 29a:261-74.
[108]    Daniel H, Jahn P, Kuntze M, Martin B. Construction and application of a $\pi\sqrt{(13)}/2$ high-resolution β-ray spectrometer connected with a tandem accelerator. Nucl Instr Meth 1970; 82: 29-44.
[109]    Tretyakov EF. Toroidal beta-spectrometer with high resolving power. Izv Akad Nauk SSSR Ser Fiz 1975; 39: 583-90. Bull Acad Sci USSR Phys Ser 1975; 39: 102-9.
[110]    Robertson RGH, Bowles TJ, Stephenson Jr GJ, Wark DL, Wilkerson JF, Knapp DA. Limit on $\overline{\nu}_e$ mass from observation of the β decay of molecular tritium. Phys Rev Lett 1991; 67: 957-60.
[111]    Holzschuh E, Kündig W, Palermo L, Stüssi H, Wenk P. A β-spectrometer for searching effects of finite neutrino mass. Nucl Instr Meth Phys Res A 1999; 423: 52-67.
[112]    Stoeffl W, Decman DJ. Anomalous structure in the beta decay of gaseous molecular tritium. Phys Rev Lett 1995; 75: 3237-40.
[113]    Tretyakov EF, Myasoedov NF, Apalikov AM, Konyaev VF, Lyubimov VA, Novikov EG. Measurement of tritium beta spectrum with the aim to improve the upper limit of the antineutrino rest mass. Izv Akad Nauk SSSR Ser Fiz 1976; 40: 2026-35. Bull Acad Sci USSR Phys Ser 1976; 40: 1-9.
[114]    Simpson JJ. Measurement of the β-energy spectrum of $^3$H to determine the antineutrino mass. Phys Rev D 1981; 23: 649-62.
[115]    Lubimov VA, Novikov EG, Nozik VZ, Tretyakov EF, Kosik VS. An estimate of the $\nu_e$ mass from the β-spectrum of tritium in the valine molecule. Phys Lett B 1980; 94: 226-68.
[116]    Lubimov VA, Novikov EG, Nozik VZ, Tretyakov EF, Kosik VS, Myasoedov NF. An estimate of the neutrino rest mass from measurement of the tritium β-spectrum (*in Russian*). Report ITEF-72 (56pp). Institute of theoretical and experimental physics. Moscow, 1981.
[117]    Boris S, Golutvin A, Laptin L, *et al*. Neutrino mass from the beta spectrum in the decay of tritium. Phys Rev Lett 1987; 58: 2019-22. Erratum in Phys Rev Lett 1988; 61: 245.
[118]    Lippmaa E, Pikver R, Suurmaa E, *et al*. Precise $^3$H-$^3$He mass difference for neutrino mass determination. Phys Rev Lett 1985; 54: 285-8.
[119]    Nagy Sz, Fritioff T, Björkhage M, Bergström I, Schuch R. On the Q-value of the tritium β-decay. Europhys Lett 2006; 74: 404-10.
[120]    Wilkerson JF. Direct searches for neutrino mass. Nucl Phys B (Proc Suppl) 1993; 31: 32-41.
[121]    Kündig W, Holzschuh E. Electron antineutrino mass from β-decay. Prog Part Nucl Phys 1994; 32: 131-51.
[122]    Kawakami H, Kato S, Ohshima T, *et al*. New upper bound on the electron anti-neutrino mass. Phys Lett B 1991; 256: 105-11.
[123]    Holzschuh E, Fritschi M, Kündig W. Measurement of the electron neutrino mass from tritium β-decay. Phys Let B 1992; 287: 381-8.
[124]    Kündig W. New neutrino mass limits. CERN Courier 1986; June iss: 15.
[125]    Lobashev VM, Spivak PE. A method for measuring the electron antineutrino rest mass. Nucl Inst Meth A 1985; 240: 305-10.
[126]    Balashov SN, Belesev AI, Bleile AI, *et al*. Integrating electrostatic spectrometer of low-energy electrons with magnetic adiabatic collimation for the measurement of the electron antineutrino rest mass (*in Russian*). Report P-0617 (18pp). Institute of nuclear research, Moscow, 1989.
[127]    Picard A, Backe H, Barth H, *et al*. A solenoid retarding spectrometer with high resolution and transmission for keV electrons. Nucl Inst Meth B 1992; 63: 345-50.
[128]    Weinheimer C, Przyrembel M, Backe H, *et al*. Improved limit on the electron-antineutrino mass from tritium β-decay. Phys Lett B 1993; 300: 210-16.
[129]    Sun Hancheng, Liang Dongqui, Chen Shiping *et al*. An upper limit for the electron anti-neutrino mass. Chinese J Nucl Phys 1993; 15: 261-8.
[130]    Feldman GJ, Cousin RD. Unified approach to the classical statistical analysis of small signals. Phys Rev D 1999; 57: 3873-89.
[131]    Saenz A, Jonsell S, Froelich P. Improved molecular final-state distribution of HeT$^+$ for the β-decay process of T$_2$. Phys Rev Lett 2000; 84: 242-5.
[132]    Lobashev VM. The search for the neutrino mass by direct method in the tritium decay and perspectives of study in the project KATRIN. Nucl Phys A 2003; 719: 153c-60c.
[133]    Babutzka M, Bahr M, Bonn J, *et al*. Monitoring of the operating parameters of the KATRIN windowless gaseous tritium source. New J Phys 2012; 14: 103046, 1-29. arXiv: 1205.5421.
[134]    Grohmann S, Bode T, Hötzel M, Schön H, Süßer M, Wahl T. The thermal behavior of the tritium source in KATRIN. Cryogenics 2013; 55-56: 5-11.
[135]    Fischer S, Sturm M, Schlösser M, *et al*. Monitoring of tritium purity during long-term circulation in the KATRIN test experiment LOOPINO using laser Raman spectroscopy. arXiv: 1208.1605 (6pp).
[136]    Prall M, Renscheler P, Glück F, *et al*. The KATRIN pre-spectrometer at reduced filter energy. New J Phys 2012; 14: 073054, 1-22. arXiv: 1203.2444.

[137] Amsbaugh F, Barrett J, Beglarian A, *et al*. Focal-plane detector system for the KATRIN experiment. Nucl Instr Meth Phys Res A 2015; 778: 40-60, 1-28. arXiv: 1404.4469.

[138] Bauer S, Berendes R, Hochschulz F, *et al*. Next generation KATRIN high precision voltage divider for voltages up to 65kV. JINST 2013; 8: 10026, 1-16. arXiv: 1309.4955.

[139] Zbořil M, Bauer S, Beck M, *et al*. Ultra-stable implanted $^{83}$Rb/$^{83m}$Kr electron sources for the energy scale monitoring in the KATRIN experiment. JINST 2013; 8: P03009, 1-30. arXiv: 1212.4955.

[140] Otten E. T$_2$-beta-spectroscopy at KATRIN and the challenge of controlling the electrostatic potentials. Nucl Phys B (Proc Suppl) 2013; 237-8: 57-60.

[141] Steinbrink N, Hannen V, Martin EL, Robertson RGH, Zacher M, Weinheimer C. Neutrino mass sensitivity by MAC-E-Filter based time-of-flight spectroscopy with the example of KATRIN. New J Phys 2013; 15:113020, 1-31. arXiv: 1308.0532.

[142] Angrik J, Armbrust T, Beglarian A, *et al*. KATRIN design report 2004. FZKA Scientific Report 7090 (245pp), Karlsruhe 2005.

[143] Sisti M, Arnaboldi C, Broffeio C, *et al*. New limits from the Milano neutrino mass experiment with thermal microcalorimeters. Nucl Inst Meth A2004; 520: 125-31.

[144] Gatti F, Fortanelli F, Galeazzi M, Swift AM, Vitale S. Detection of environmental fine structure in the low-energy β-decay spectrum of $^{187}$Re. Nature 1999; 397: 137-9.

[145] Gatti F, Gallinaro G, Pergolessi D, *et al*. MARE − Microcalorimetric array for a rhenium experiment. 2006 (149pp), http://crio.mib.infn.it/wig/silicini/proposal/proposal_MARE_v2.6.pdf .

[146] Nucciotti A. Neutrino mass calorimetric searches in the MARE experiment. Nucl Phys Proc Suppl 2012; 229-232: 155-9, arXiv:1012.2290.

[147] Ferri E. The status of the MARE experiment with $^{187}$Re and $^{163}$Ho isotopes. The 13$^{th}$ Int. conf. on topics in astroparticle and underground physics, Asilomar, California USA, Sept 8-13, 2013.
*https://conferences.lbl.gov/event/36/session/17/contribution/157/material/slides/0.pdf*

[148] DeRújula A, Lusignoli M. Calorimetric measurements of $^{163}$Ho decay as tools to determine the electron neutrino mass. Phys Lett B 1982; 118: 429-34.

[149] De Rújula A. Two old ways to measure the electron-neutrino mass. arXiv: 1305.4857 (11pp).

[150] Springer PT, Bennett CL, Baisden PA. Measurement of the neutrino mass using the inner bremsstrahlung emitted in the electron-capture decay of $^{163}$Ho. Phys Rev A 1987; 35: 679-89.

[151] Nucciotti A. Statistical sensitivity of $^{163}$Ho electron capture neutrino mass experiments. Eur Phys J C 2014; 74: 3161, 1-6. arXiv: 1405.5060.

[152] Alpert B, Balata M, Bennett D, *et al*. HOLMES − the electron capture decay of $^{163}$Ho to measure the electron neutrino mass with sub-eV sensitivity. Eur Phys J C 2015 (in press). arXiv: 1412.5060 (11pp).

[153] Robertson RGH. Can neutrino mass be measured in low-energy electron capture decay? arXiv: 1411.2906 (12pp).

[154] Faessler A. Improved description of one- and two-hole excitations after electron capture in $^{163}$Ho and the determination of the neutrino mass. arXiv: 1501.04338 (17pp).

[155] Faessler A, Enss C, Gastaldo L, F. Šimkovic F. Determination of the neutrino mass by electron capture in 163 Holmium and the role of the three-hole states in 163 Dysprosium. arXiv:1503.02282 (13pp).

[156] Gastaldo L, Blaum K, Doerr A, *et al*. The Electron Capture 163Ho Experiment ECHo. J Low Temp Phys 2014; 176: 876-84.

[157] Ranitzsch PC-O, Porst J-P, Kempf S, *et al*. Development of magnetic calorimeters for high precision measurements of calorimetric $^{187}$Re and $^{163}$Ho spectra. J Low Temp Phys 2012; 167: 1004-14.

[158] Blaum K, Doerr A, Duellmann CE, *et al*. The electron capture $^{163}$Ho experiment ECHo. arXiv: 1306.2625 (6pp).

[159] Gastaldo L. $^{163}$Ho based experiments. AIP Conf Proc 2015; 1666: 05001, 1-6.

[160] Monreal B, Formaggio JA. Relativistic cyclotron radiation detection of tritium decay electrons as a new technique for measuring the neutrino mass. Phys Rev D 2009; 80: 051301(R), 1-4. arXiv: 0904.2860.

[161] Doe PJ, Kofron J, McBride EL, *et al*. Project 8: Determining neutrino mass from tritium beta decay using a frequency-based method. arXiv: 1309.7093 (8pp).

[162] Asner DM, Bradley RF, de Viveiros L, *et al*. Single electron detection and spectroscopy via relativistic cyclotron radiation. arXiv: 1408.5362 (6pp).

[163] Cohen SG, Murnick DFE, Raghavan RS. Bound-state beta-decay and kinematic search for neutrino mass. Hyperfine Interact 1987; 33: 1-8.

[164] Jung M, Bosch F, Beckert K, *et al*. First observation of bound-state β$^−$ decay. Phys Rev Lett 1992; 69: 2164-7.

[165] Bahcall JN. Theory of bound-state beta decay. Phys Rev 1961; 124: 495-9.

[166] Lindroos M, McElrath B, Orme C, Schwetz T. Measuring neutrino mass with radioactive ions in a storing ring. Eur Phys J C 2009; 64: 549-60. arXiv: 0904.1089.

[167] Kopp J, Merle A. Ultralow Q values for neutrino mass measurements. Phys Rev C 2010; 81: 045501,1-5. arXiv: 0911.3329.

[168] Jerkins M, Klein JR, Majors JH, Robicheaux F, Raizen MG. Using cold atoms to measure neutrino mass. New J Phys 2010; 12: 043022, 1-9.

[169] Otten E. Comment on "Using cold atoms to measure neutrino mass". New J Phys 2011; 13: 078001, 1-5.

[170] Jerkins M, Klein JR, Majors JH, Robicheaux F, Raizen MG. Reply to comment on "Using cold atoms to measure neutrino mass". New J Phys 2011; 12: 078002, 1-4.

[171] Mårtensson N, Baltzer P, Brühwiller PA, *et al*. A very high resolution electron spectrometer. J Electron Spec Relat Phenom 1994; 70: 117-28.

[172] Fukumi A, Kuma S, Miyamoto Yu, *et al*. Neutrino spectroscopy with atoms and molecules. Prog Theor Exp Phys 2012; 04D002, 1-79. arXiv: 1211.4904.

[173] Dinh DN, Petcov ST, Sasao N, Tanaka M, Yoshimura M. Observables in neutrino mass spectroscopy using atoms. Phys Lett B 2013; 719: 154-63.

[174] Tashiro M, Ehara M, Kuma S, *et al*. Iodine molecule for neutrino mass spectroscopy: ab initio calculation of spectral rate. Prog Theor Exp Phys 2014; 013B02, 1-21. arXiv: 1310.7342.
[175] Yoshimura M, Sasao N, Tanaka M. Dynamics of two-photon paired superradiance. Phys Rev A 2012; 86: 013812, 1-14. arXiv: Abe K, Haga Y, Hayoto Y, *et al*. Limits on sterile neutrino mixing using atmospheric neutrinos in Super-Kamiokande. arXiv: 1410.2008 (23pp).
[176] Dinh DN, Petcov ST. Radiative emission of neutrino pairs in atoms and light sterile neutrinos. arXiv: 1411.7459 (18pp).
[177] Kayser B. Are there sterile neutrinos? arXiv: 1402.3028 (3pp).
[178] Lasserre T. Light sterile neutrinos in particle physics: experimental status. Phys Dark Univ 2014;4: 81-5. arXiv: 1404.7352.
[179] Archidiacono M, Fornengo N, Gariazzo S, Giunti C, Hannestad S, Laveder M. Light sterile neutrinos after BICEP-2. J Cosm Astropart Phys 2014; 06: 031, 1-11. arXiv: 1404.1794.
[180] Gariazzo S, Giunti C, Laveder M. Light sterile neutrinos in cosmology and short-baseline oscillation experiments. J High Ener Phys 2013; 211, 1-12. arXiv: 1309.3192.
[181] Spinelli M. Cosmological constraints on neutrinos with Planck data. AIP Conf Proc 2015; 1666: 14001, 1-7.
[182] An FP, Balantekin AB, Band HR, *et al*. Search for a light sterile neutrino at Daya Bay. arXiv: 1407.7259 (7pp).
[183] Abe K, Haga Y, Hayoto Y, *et al*. Limits on sterile neutrino mixing using atmospheric neutrinos in Super-Kamiokande. arXiv: 1410.2008 (23pp).
[184] Boyarsky A, Iakubovskyi D, Ruchayskiy O. Next decade of sterile neutrino studies. Phys Dark Univ 2012; 1: 136-54. arXiv: 1306.4954.
[185] Leistedt B, Peiris HV, Verde L. No new cosmological concordance with massive sterile neutrinos. Phys Rev Lett 2014; 113: 041301, 1-6. arXiv: 1404.5950.
[186] Schreckenbach K, Colvin G, von Feilitzsch F. Search for mixing of heavy neutrinos in the $\beta^+$ and $\beta^-$ spectra of the $^{64}$Cu decay. Phys Lett B 1983; 129: 265-8.
[187] Mampe W, Schreckenbach K, Jeuch P, *et al*. The double focusing iron-core electron spectrometer "BILL" for high resolution (n, e$^-$) measurements at the high flux reactor in Grenoble. Nucl Inst Meth 1978; 154: 127-149.
[188] Van Elmbt L, Deutsch J, Prieels R. Measurement of the $^{20}$F beta spectrum: A low-energy test of the standard electro-weak gauge-model. Nucl Phys A 1987; 469: 531-56.
[189] Deutsch J, Lebrun M, Prieels R. Searches for admixture of massive neutrinos into the electron flavor. Nucl Phys A 1990; 518: 149-55.
[190] Simpson JJ. Evidence of heavy-neutrino emission in beta decay. Phys Rev Lett 1985; 54: 1891-3.
[191] Glashow SL. A novel neutrino mass hierarchy. Phys Lett B 1991; 256: 255-7.
[192] Hime A, Jelley NA. New evidence for the 17 keV neutrino. Phys Lett B 1991; 257: 441-9.
[193] Ohshima T, Sakamoto H, Sato T, *et al*. No 17 kev neutrino: Admixture < 0.073% (95% CL). Phys Rev D 1993; 47: 4840-56.
[194] Mortara JL, Ahmad I, Coulter KP, *et al*. Evidence against a 17 keV neutrino from $^{35}$S beta decay. Phys Rev Lett 1993; 70: 394-7.
[195] Bowler MG, Jelley NA. Investigations into the origin of the spurious 17 keV neutrino signal observed in $^{35}$S beta decay. Z Phys C 1995; 68: 391-414.
[196] Simpson JJ, Hime A. Evidence of the 17-keV neutrino in the β spectrum of $^{35}$S. Phys Rev D 1989; 39: 1825-36.
[197] Müller S, Chen Shiping, Daniel H, *et al*. Search for an admixture of a 17 keV neutrino in the β decay of $^{35}$S. Z Naturforsch 1994; 49a: 874-84.
[198] Montanet L, Barnett RM, Groom DE, *et al*. Review of particle properties. Phys Rev D 1994; 50: 1173-1826.
[199] Caso C, Conforto G, Gurter A, *et al*. Reviev of particle properties. Eur Phys J C 1998; 3: 1-783.
[200] Wietfeldt FE, Norman EB. The 17 keV neutrino. Phys Rep 1996; 273: 149-97.
[201] Franklin A. The appearance and disappearance of the 17-keV neutrino. Rev Mod Phys 1995; 67: 457-90.
[202] Morrison DRO. The rise and fall of the 17-keV neutrino. Nature 1993; 366: 29-32.
[203] Hiddemann K-H, Daniel H, Schwentker O. Limits on neutrino masses from the tritium β spectrum. J Phys G: Nucl Part Phys 1995; 21: 639-50.
[204] Holzschuh E, Kündig W, Palermo L, Stüssi H, Wenk P. Search for heavy neutrinos in the β-spectrum of $^{63}$Ni. Phys Lett B 1999; 451: 247-55.
[205] Holzschuh E, Palermo L, Stüssi H, Wenk P. The β-spectrum of $^{35}$S and search for the admixture of heavy neutrinos. Phys Lett B 2000; 482: 1-9.
[206] Galeazzi M, Fortanelli F, Gatti F, Vitale S. Limits on the existence of heavy neutrinos in the range 50–1000 eV from the study of the $^{187}$Re beta spectrum. Phys Rev Lett 2001; 86: 1978-81.
[207] Bonvicini G. Statistical issues in the 17-keV neutrino experiments. Z Phys A 1993; 345: 97-117.
[208] Apalikov AM, Boris SD, Golutvin AI, *et al*. Search for heavy neutrinos in β decay. Pis'ma Zh Eksp Teor Fiz 1985; 42: 233-6. JETP Lett 1985; 42: 289-93.
[209] Dragoun O, Špalek A, Ryšavý M, *et al*. Search for an admixture of heavy neutrinos in the β-decay of $^{241}$Pu. J Phys G: Nucl Part Phys 1999; 25: 1839-58.
[210] Varga D, Kádár I, Kövér Á, et al. Electrostatic spectrometer for measurement of internal conversion electrons in the 0.1 – 20 keV region.Nucl Inst Meth 1982; 192: 277-86.
[211] Briançon Ch, Legrand B, Walen RJ, Vylov Ts, Minkova A, Inoyatov A. A new combined electrostatic electron spectrometer. Nucl Instr Meth Phys Res 1984; 221: 547-57.
[212] Campbell JL, Papp T. Width of the atomic K – N$_7$ levels. Atom Data Nucl Data Tabl 2001; 77: 1-56.
[213] Long X, Liu M, Ho F, Peng X. Cross sections for *K*-shell ionization by electron impact. Atom Data Nucl Data Tables 1990; 45: 353-66.
[214] Dragoun O, Špalek A, Ryšavý M. Improved methods of measurement and analysis of conversion electron and β-particle spectra. Appl Radiat Isot 2000; 52: 387-91.

[215] Dragoun O, Ryšavý M, Dragounová N, Brabec V, Fišer M. An improved method for the measurement of fine effects in electron spectra. Nucl Instr Meth Phys Res A 1995; 365: 385-91.

[216] Dragoun O, Ryšavý M, Špalek A, *et al*. Statistical tests of invariability of the measurement conditions in the β-ray spectroscopy. Nucl Instr Meth Phys Res A 1997; 391: 345-50.

[217] Belesev AI, Berlev AI, Geraskin EV, *et al*. An upper limit on additional neutrino mass eigenstate in 2 to 100 eV region from "Troitsk nu-mass" data. Pis'ma v ZhETF 2013; 97: 73-5, arXiv: 1211.7193.

[218] Belesev AI, Berlev AI, Geraskin EV, *et al*. A search for an additional neutrino mass eigenstate in 2 to 100 eV region from "Troitsk nu-mass"data–detailed analysis. J Phys G Nucl Part Phys 2014; 41: 015001,1-13. arXiv: 1307.5687.

[219] Kraus C, Singer A, Valerius K, Weinheimer C. Limit on sterile neutrino contribution from the Mainz neutrino mass experiment. Eur J Phys C 2013; 73: 2323, 1-14. arXiv: 1210.4194.

[220] Formaggio JA, Barrett J. Resolving the reactor neutrino anomaly with the KATRIN neutrino experiment. Phys Lett B 2011; 706: 68-71. arXiv: 1105.1326.

[221] Mention G, Fechner M, Lasserre Th, *et al*. Reactor antineutrino anomaly. Phys Rev D 2011; 83: 073006, 1-20. arXiv: 1101.2755.

[222] Zhang C, Quian X, Vogel P. Reactor antineutrino anomaly with known $\theta_{13}$. Phys Rev D 2013; 87: 073018, arXiv: 1303.0900.

[223] Garvey GT, Hayes AC, Jungman G, Jonkmans G. Further investigation of the „reactor anomaly". AIP Conf Proc 2015; 1666: 080001, 1-7.

[224] Esmaili A, Peres OLG. KATRIN sensitivity to sterile neutrino mass in the shadow of lightest neutrino mass. Phys Rev D 2012; 85: 117301, 1-5. arXiv: 1203.2632.

[225] Sejersen Riis A, Hannestad S. Detecting sterile neutrinos with KATRIN like experiments. J Cosm Astropart Phys 2011; 02: 011, 1-23. arXiv: 1008.1495.

[226] Bonn J, Bornschein L, Degen B, Otten EW, Weinheimer Ch. A high resolution electrostatic time-of-flight spectrometer with adiabatic magnetic collimation. Nucl Instr Meth Phys Res A 1999; 421: 256-65.

[227] Mertens S, Dolde K, Korzeczek M, *et al*. Wavelet approach to search for sterile neutrinos in tritium β-decay spectra. Phys. Rev. D 2015; 91: 042005, 1-9. arXiv: 1410.7684.

[228] Filianin PE, Blaum K, Eliseev SA, *et al*. On the keV sterile neutrino search in electron capture. J Phys G 2014; 41: 095004, 1-16. arXiv: 1402.4400.

[229] Abdurashitov DN, Berlev AI, Likhovid NA, Lokhov AV, Tkachev II, Yants VE. Search for an admixture of sterile neutrinos via registering tritium decays in a proportional counter: new possibilities (in Russian). arXiv: 1403.2935 (22pp).

[230] Kalbfleisch GR, Bahran MY. Experimental limits on heavy neutrinos in tritium beta decay. Phys Lett B 1993; 303: 355-8.

[231] Abdurashitov DN, Belesev AI, Berlev AI, *et al.* The current status of "Troitsk nu-mass" experiment in search for sterile neutrino. arXiv: 1504.00544 (17pp).

[232] Shi X, Fuller GM. New Dark Matter Candidate: Nonthermal Sterile Neutrinos. Phys Rev Lett 1999; 82: 2832-5.

[233] de Vega HJ, Moreno O, Moya de Guerra E, Ramón edrano , Sánchez NG. Role od sterile neutrino warm dark matter in rhenium and tritium beta decays. Nucl Phys B 2013; 866: 177-195.

[234] Barry J, Heeck J, RodejoAhann W. Sterile neutrinos and right-handed currents in KATRIN. J High Energy Phys 2014; 07: 081. arXiv: 1404.5955 (20pp).

[235] Rodejohann W, Zhang H. Signatures of extra dimensional sterile neutrinos. arXiv: 1407.2739 (12pp).